\documentclass[aps,prl,twocolumn,superscriptaddress,nofootinbib,floatfix]{revtex4-2}

\usepackage{amsmath,amssymb,bm}
\usepackage{graphicx}
\usepackage{booktabs}
\usepackage{xcolor}
\usepackage[
    colorlinks=true,
    citecolor=blue,
    linkcolor=black,
    urlcolor=blue
]{hyperref}
\usepackage{ulem}

\newcommand{\ii}{\mathrm{i}}
\newcommand{\Tr}{\mathrm{Tr}}

\begin{document}
\title{Probing Non-Abelian Topology through Quench Dynamics}

 \author{Zhenming Zhang}
 \affiliation{Laboratory of Quantum Information, University of Science and Technology of China, Hefei 230026, China}
 \affiliation{Hefei National Laboratory, University of Science and Technology of China, Hefei 230088, China}
 \author{Tianyu Li}
\email{tianyuli@m.scnu.edu.cn}
\affiliation {Key Laboratory of Atomic and Subatomic Structure and Quantum Control (Ministry of Education), Guangdong Basic Research Center of Excellence for Structure and Fundamental Interactions of Matter, School of Physics, South China Normal University, Guangzhou 510006, China}
\affiliation {Guangdong Provincial Key Laboratory of Quantum Engineering and Quantum Materials, Guangdong-Hong Kong Joint Laboratory of Quantum Matter, Frontier Research Institute for Physics, South China Normal University, Guangzhou  510006, China}
\author{Wei Yi}
\email{wyiz@ustc.edu.cn}
\affiliation{Laboratory of Quantum Information, University of Science and Technology of China, Hefei 230026, China}
\affiliation{Anhui Province Key Laboratory of Quantum Network, University of Science and Technology of China, Hefei 230026, China}
\affiliation{CAS Center For Excellence in Quantum Information and Quantum Physics, Hefei 230026, China}
\affiliation{Hefei National Laboratory, University of Science and Technology of China, Hefei 230088, China}

\begin{abstract}
Quench dynamics offer valuable insights to topological matter, wherein the emergent topological phenomena serve as a dynamic probe to the topology of static Hamiltonians. 
While such a practice has been widely applied to detect the Abelian band topology or the Euler-class topology of band nodes, little is known regarding the quench dynamics of non-Abelian topological insulators in multiband systems, where the intertwining geometry of all eigenstates necessitates the description of noncommutative topological charges.
Here we propose a quench protocol for non-Abelian topological insulators, so that 
the underlying topology can be systematically probed using dynamic signatures.
Specifically, we develop a band-projection scheme to extract the gap-resolved topology from dynamic skyrmions in the emergent quasimomentum--time manifolds of the quench processes.
These skyrmions are characterized by a series of gap-resolved dynamic Chern numbers, which enable us to reconstruct the full non-Abelian topology of the post-quench Hamiltonian, including the global non-Abelian topological charge and edge-state patterns under the open-boundary condition.
Our proposal offers dynamical access to non-Abelian topology in multiband systems, and 
can be readily implemented using photons or cold atoms.
\end{abstract}

\maketitle

{\it Introduction.---}
Non-Abelian topological insulator, a latest addition to topological matter, arises in systems with multiple band gaps, wherein the geometry of eigenstates in the Hilbert space is captured only by noncommutative topological charges~\cite{wu2019nonabelian,guo2021charges,yang2024nonabelian}.
As they are closely related to non-Abelian phenomena such as the braiding and linking of band nodes~\cite{unal2020euler,zhao2022euler,bouhon2020braiding,tiwari2020nodal,yang2020links,peng2022phonons,wang2022nonadjacent,wang2022earring,wang2023frameflow,breach2024interferometry} and periodically driven Floquet non-Abelian phases~\cite{li2023floquet,slager2024floquet,lin2026floquet,qiu2026floquet,hu2026scattering}, the recent experimental demonstration of these phenomena in synthetic platforms have attracted much attention~\cite{guo2021charges,lin2026floquet,qiu2026floquet,hu2026scattering,jiang2021fourband,jiang2026sixband,jiang2021acoustic,qiu2023braiding,wang2025braiding}.
Nevertheless, the direct visualization and detection of the non-Abelian topological charges remains challenging.

In conventional topological insulators, quantum quenches have proved to be a powerful tool in probing the underlying topological invariant~\cite{caio2015chern,hauke2014tomography,chang2018topology,gong2018entanglement,zhang2018classification,hu2020quench}, as the appearance of topological phenomena such as the dynamic quantum phase transitions~\cite{vajna2015dpt,budich2016order,flaschner2018vortices}, or emergent links~\cite{wang2017chern,tarnowski2019linking}, rings~\cite{sun2018quench}, and skyrmions~\cite{yang2018dynamical,wang2019skyrmions,qiu2019fixed}, are dictated by the band topologies of the pre- and post-quench Hamiltonians. 
However, previous quench schemes are either based on two-band models with a single band gap~\cite{wang2017chern,tarnowski2019linking,sun2018quench,yang2018dynamical,wang2019skyrmions,qiu2019fixed},
or focused on the Euler-class topology associated with band nodes for multi-band-gap systems in two dimensions~\cite{unal2020euler,zhao2022euler}.
It thus remains elusive as to the feasibility of dynamically probing the band topology of one-dimensional non-Abelian insulators that are fully gapped.

In this work, we propose a gap-resolved quench protocol for non-Abelian topological insulators with three bands, such that the non-Abelian topology of the system can be probed from emergent topological phenomena in the quench dynamics. 
Our protocol involves a series of quenches from different initial states, governed by the same post-quench Hamiltonian, so that the non-Abelian topology of the post-quench Hamiltonian can be reconstructed combining the state evolution in all quench processes.
Key to our proposal is a band-projection scheme, which enables the visualization, on the Bloch sphere, of the relative dynamics between bands below and above a given gap. The gap-resolved topology then manifests itself as dynamic skyrmions in the spin texture across the emergent quasimomentum--time manifold. These skyrmions are characterized by 
gap-resolved Chern numbers, whose magnitudes and signs combined fully determine the non-Abelian topology of the post-quench Hamiltonian. 
We show how the dynamic signatures can help detect topological phase transitions, and discuss the extension of our scheme to general multiband cases.

{\it Non-Abelian topological insulator.---}
We consider the one-dimensional three-band tight-binding Hamiltonian~\cite{guo2021charges}
\begin{equation}\label{eq:real-space-hamiltonian}
\begin{aligned}
  H
  =&\sum_{n,\alpha}s_{\alpha\alpha}
  c^\dagger_{\alpha,n}c_{\alpha,n} \\
  &+\sum_{n,\alpha,\beta}
  \left(
  v_{\alpha\beta}c^\dagger_{\alpha,n}c_{\beta,n+1}
  +v'_{\alpha\beta}c^\dagger_{\alpha,n}c_{\beta,n+2}
  +{\rm H.c.}
  \right),
\end{aligned}
\end{equation}
where $c_{\alpha,n}$ ($c^\dag_{\alpha,n}$) is the annihilation (creation) operator on sublattice site $\alpha$ ($\alpha\in \{A,B,C\}$) of the $n$th unit cell. 
The on-site potentials $s_{\alpha\alpha}$ are taken to be real, while the inter- and intra-cell hopping coefficients satisfy $v_{AB}=\ii u$, $v_{BC}=\ii v$, $v_{AC}=\ii w$, $v'_{AB}=\ii u'$, $v'_{BC}=\ii v'$, and $v'_{AC}=\ii w'$, where $u,v,w,u',v',w'$ are all real.

\begin{figure}[t]
  \centering
  \includegraphics[width=\columnwidth]{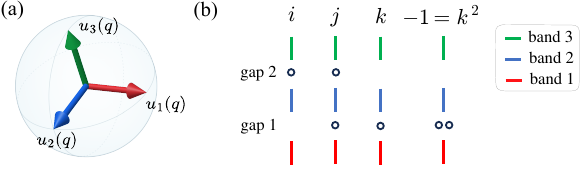}
\caption{
(a) The orthonormal real eigenvectors $u_{1,2,3}(q)$ of the PT-symmetric
Hamiltonian $H(q)$ form an eigenframe
$F(q)=[u_1(q),u_2(q),u_3(q)]\in O(3)$,
where $q$ denotes quasimomentum.
(b) Schematic illustration for typical quarternion charges and edge-mode patterns. 
The colored vertical lines denote the energy bands, and each circle represents a pair of edge modes under the open boundary condition. The bands and gaps are labeled in ascending order according to the eignenergies.
For the $-1$ case, only the realization $-1=k^2 $ is shown,
where two pairs of edge modes exist in gap 1.
}
  \label{fig1}
\end{figure}

With these, Hamiltonian (\ref{eq:real-space-hamiltonian}) is parity-time (PT) symmetric~\cite{supplemental}. It follows that the corresponding Bloch Hamiltonian $H(q)$ is completely real for any quasimomentum $q$, and its eigenstates can be chosen as real vectors
$u_1(q)$, $u_2(q)$, and $u_3(q)$, forming an orthonormal eigenframe 
$
F(q)=\big[u_1(q),u_2(q),u_3(q)\big]\in O(3)
$ [see Fig.~\ref{fig1}(a)], 
where $O(N)$ denotes the group of $N\times N$ real orthogonal matrices.
Here the eigenvectors are labeled in ascending order of their eigenenergies, consistent with the band-gap labels [see Fig.~\ref{fig1}(b)]. 
Furthermore, as the overall sign of each eigenvector is physically irrelevant, the
physical eigenframe space is
$
M_3=O(3)/(\pm1)^3
$, and the band topology of the system is given by
$
\pi_1(M_3)=Q_8=\{\pm1,\pm i,\pm j,\pm k\}
$, 
where $Q_8$ is the quaternion group~\cite{wu2019nonabelian,guo2021charges,bouhon2020geometry}. 
The group elements of $Q_8$ are quaternion charges satisfying $i^2=j^2=k^2=ijk=-1$, 
and the charges $i,j,k$ anticommute with one another, underlining the global non-Abelian topology.

Geometrically, the quaternion charges characterize the eigenframe rotations
as the quasimomentum traverses the first Brillouin zone (BZ)~\cite{guo2021charges,wang2023frameflow,jiang2022twod}.
For example, charge $k$ corresponds to a $\pi$ rotation of the eigenframe
with respect to $u_3$, generating a nontrivial Zak phase of $\pi$ for gap $1$ (defined as the sum of the Zak phases, modulo $2\pi$, of all bands below the said gap)~\cite{supplemental,zak1989berry,ahn2019stiefel}.
Similarly, charge $i$ generates a Zak phase of $\pi$ for gap $2$.
Under the open boundary condition, these would each give rise to a pair
of topological edge modes within the corresponding gap
[see Fig.~\ref{fig1}(b)]~\cite{guo2021charges}.
More interestingly, charge $j$ corresponds to a $\pi$ rotation of
the eigenframe around the $u_2$ axis, under which both gaps acquire
nontrivial Zak phases of $\pi$, anticipating a pair of edge modes within each gap (under the
open boundary condition).
This is consistent with the decomposition $j=ki$, which makes explicit the individual contributions from the two gaps~\cite{zhang2026topoword}.
More generally, as $\pm k$ and $\pm i$ generate the quaternion group $Q_8$, any quaternion charge can be decomposed as their ordered product.
The utility of such a gap-resolved topology is further underscored by 
the case of charge $-1$, where different edge-mode configurations motivate
distinct decompositions with the corresponding gap-resolved topology~\cite{guo2021charges,jiang2022twod,zhang2026topoword},
For example, one of the realizations hosts two pairs of edge modes in gap 1, wherein the global topological charge can be
decomposed as $-1=k^2$, for a direct correspondence.

\begin{figure}[t]
  \centering
  \includegraphics[width=\columnwidth]{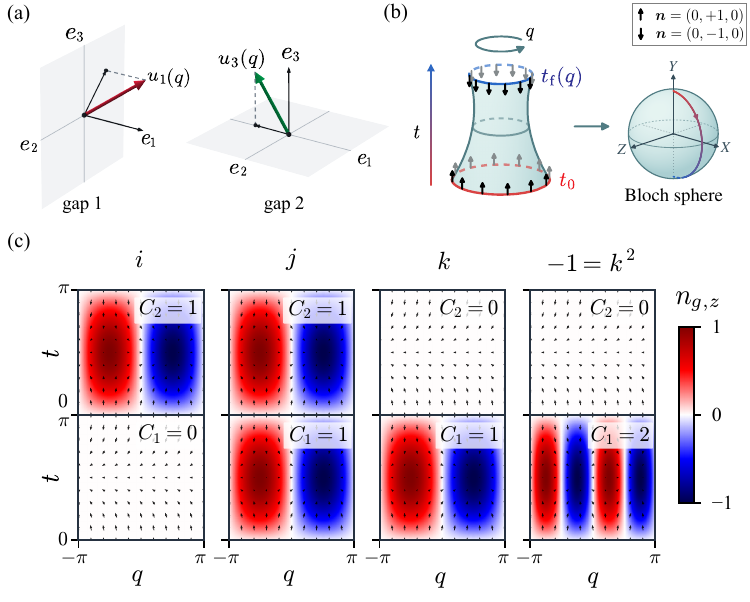}
\caption{
(a) Basis construction for gap-resolved evolution operators,
with $e_i=u_i(q_0)$ fixed at quasimomentum $q_0$. For gaps 1 and 2, we retain
$e_1$ and $e_3$, respectively, while the second basis vectors are given by the
normalized projection of $u_1(q)$ onto the $(e_2,e_3)$ plane, and of $u_3(q)$
onto the $(e_1,e_2)$ plane, respectively. 
(b) Map from the quasimomentum--time manifold (left) to the Bloch sphere (right). The Bloch vector $\bm{n}(q,t)$ is uniform along each time boundary, $t=0$ and $t=t_{\mathrm f}(q)$, so that each can be contracted to a point, allowing the manifold to be deformed into a sphere. The trajectory on the Bloch sphere illustrates the evolution from $\bm{n}(q,0)=(0,1,0)$ to $\bm{n}(q,t_f(q))=(0,-1,0)$.
(c) Gap-resolved Bloch vectors $\bm n_2(q,t)$ (upper row) and
$\bm n_1(q,t)$ (lower row) for cases with global topological charges $i$, $j$, $k$, and $-1=k^2$, respectively.
The color bar denotes $n_{g,z}$, and the arrows $(n_{g,x},n_{g,y})$. Gap-resolved Chern numbers for each case are shown as legends.}
\label{fig2}
\end{figure}

{\it Gap-resolved quench dynamics.---}
In the conventional Abelian topological insulators, the band topology can be characterized by the rotation of a single eigenvector (for one of the two bands) across the BZ. Such an understanding underlies topological quenches, wherein the system is initialized in a local Wannier state of a given band and subsequently evolved under a post-quench Hamiltonian~\cite{wang2019skyrmions}. The band topology of this post-quench Hamiltonian can be probed through dynamic topological phenomena, often characterized by emergent topological invariants defined in the quasimomentum--time manifold, in the quench dynamic~\cite{zhang2018classification,hu2020quench,wang2017chern,yang2018dynamical}.
While it is natural to explore the possibility of topological quenches for non-Abelian topological insulators, key challenges remain.

First, as discussed above, non-Abelian topological insulators are
characterized by the global rotations of eigenvectors belonging to
multiple (more than two) bands.
Quench dynamics initialized over a single band is therefore insufficient
to capture the full non-Abelian topology.
We therefore consider separate quenches initialized, respectively,
on the three sublattice sites within a single unit cell of the bulk. The system is subsequently driven by the same post-quench Hamiltonian.
In the quasimomentum space, the eigenframe formed by these
initial states is the quasimomentum-independent identity matrix $F(q,0)=I$.
Apparently, the time-evolved frame at time $t$, assembled from
the time-evolved states as $F(q,t)=U(q,t):=\exp[-\ii H(q)t]$, 
carries full information of the post-quench Hamiltonian $H(q)$.

\begin{figure*}[t]
  \centering
  \includegraphics[width=\textwidth]{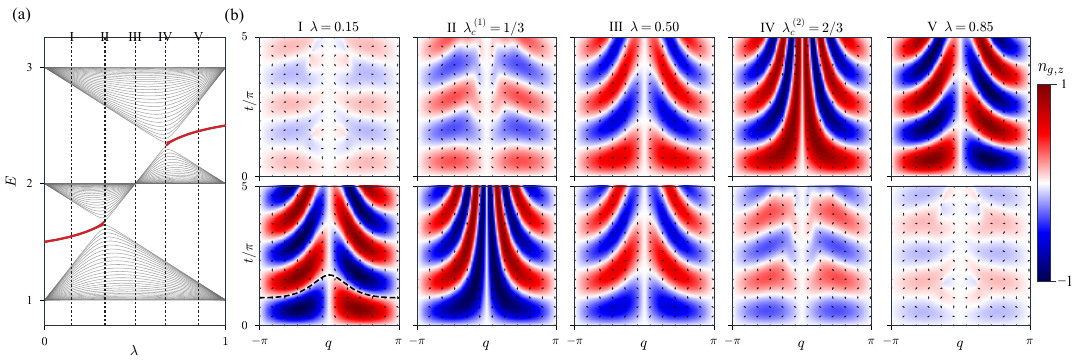}
\caption{
Open-boundary spectrum and gap-resolved Bloch vectors across topological phase
transitions.
(a) Open-boundary spectrum along the interpolation defined in
Eq.~\eqref{eq:k-to-i-path}. Gray curves represent bulk-state spectrum, while red
curves indicate those of edge modes. The five vertical dashed lines mark
the interpolation parameters used in (b). 
(b) Gap-resolved Bloch vectors of the evolution operator $U_\lambda (q,t)$. Each column is generated from a common post-quench Hamiltonian evolution, with gap 2 shown in the upper row and gap 1 in the lower row. The black dashed curve in the left-most panel of the lower row marks the quasimomentum-dependent $t_{\mathrm{f}}(q)$.}
  \label{fig3}
\end{figure*}

A more fundamental difficulty lies in the Abelian nature of higher homotopy groups $\pi_n$ for $n\geq2$, which implies that 
emergent topological invariants defined over the quasimomentum--time manifold 
are Abelian and incapable of describing the non-Abelian topology by themselves.
To solve this issue, we develop a gap-resolved quench protocol, where 
the global non-Abelian topology can be reconstructed by combining: i) all relevant emergent topological invariants in the quench dynamics, and ii) the information regarding to which energy gap each of them belongs.

Specifically, for each energy gap between adjacent bands, we study the relative dynamics between bands below and above the gap, through a projective construction of a gap-resolved evolution operator for an effective two-band model.
Taking gap $1$ for instance, 
we choose a reference quasimomentum $q_0$, and denote the eigenstates of $H(q_0)$ as $e_n=u_n(q_0)$ for $n=1,2,3$. 
The relative dynamics between band 1 and other bands is encoded in how the eigenstate $u_1(q)$
orientates itself away from $e_1$ toward the plane spanned by $e_2$ and $e_3$,
as the quasimomentum $q$ varies.
Mathematically, this information is captured by a two-dimensional basis, consisting of
$e_1$ and the projection of $u_1(q)$ onto the $e_2$--$e_3$ plane
[see Fig.~\ref{fig2}(a)]. 
This motivates us to construct the
gap-resolved evolution operator
\begin{equation}\label{eq:gap-evolutions}
  U_{1}(q,t)=W^\dagger_{1}(q)U(q,t)W_{1}(q),
\end{equation}
where 
\begin{equation}
  W_{1}(q)=
  \left(
  e_1,\,
  \chi_1(q)\frac{P_{23}u_1(q)}
  {\|P_{23}u_1(q)\|}
  \right),
  \label{eq:w-gap1}
\end{equation}
and $P_{23}=e_2e_2^T+e_3e_3^T$. Here $\chi_1(q)=\pm 1$ indicates the right- (left-) handed helical rotations about $e_1$, respectively~\cite{supplemental}.
Similarly, for gap $2$, we have
$U_{2}(q,t)=W^\dagger_{2}(q)U(q,t)W_{2}(q)$, where
\begin{equation}
  W_{2}(q)=
  \left(
  e_3,\,
  \chi_2(q)\frac{P_{12}u_3(q)}
  {\|P_{12}u_3(q)\|}
  \right),
  \label{eq:w-gap2}
\end{equation}
with $P_{12}=e_1e_1^T+e_2e_2^T$, and \(\chi_2(q)=\pm 1\) for right- (left-) handed helical rotations about $e_3$, respectively. 
Apparently, the gap-resolved dynamics driven by $U_{1,2}(q,t)$
can each be understood as a quench of an effective
two-band Hamiltonian with chiral symmetry~\cite{supplemental}.

\begin{table}[t]
\centering
\setlength{\tabcolsep}{4pt}
\begin{tabular}{ccc}
\toprule
 charge &\quad  decomposition & $(C_1,C_2)$ \\
\midrule
$+1$ & $1$  & $(0,0)$ \\
\midrule
$+i$ & $i$  & $(0,1)$ \\
$-i$ & $-i$ & $(0,-1)$ \\
\midrule
$+j$ & $ki$ & $(1,1)$ \\
$-j$ & $(-i)(-k)$ & $(-1,-1)$ \\
\midrule
$+k$ & $k$  & $(1,0)$ \\
$-k$ & $-k$ & $(-1,0)$ \\
\midrule
  & $k^2$ or $(-k)^2$
     & $(2,0)$ or $(-2,0)$ \\
$-1$ & $i^2$ or $(-i)^2$
     & $(0,2)$ or $(0,-2)$ \\
  & $(ki)^2$ or $(ik)^2$
     & $(2,2)$ or $(-2,-2)$ \\
\bottomrule
\end{tabular}
\caption{
Correspondence among the quaternion charges, their generator decompositions,
and gap-resolved Chern numbers $(C_1,C_2)$ for three-band non-Abelian topological insulators
with nearest-neighbor and next-nearest-neighbor hoppings.
The orthogonality of the eigenstates constrains the relative signs of
$C_g$; in particular, $C_1$ and $C_2$ always have the same sign
when the global charge is $\pm j$~\cite{supplemental}.
The charge $-1$ admits three distinct gap-resolved realizations, with different gap-resolved Chern numbers.
}
\label{tab:q8-skyrmion}
\end{table}

To visualize the gap-resolved dynamics on the Bloch sphere, we introduce the normalized Bloch vector
$\bm{n}_g=(n_{g,x},n_{g,y},n_{g,z})/
[{n_{g,x}^2+n_{g,y}^2+n_{g,z}^2}]^{1/2}$,
with $n_{g,\mu}$ given by
\begin{equation}
  n_{g,\mu}(q,t)=\Tr[
  U_g(q,t)\sigma_yU_g^\dagger(q,t)\sigma_\mu],
\label{eq:gap-bloch-vector}
\end{equation}
where $\sigma_\mu$ ($\mu=x,y,z$) are the Pauli matrices, and the subscript $g$ is the gap index.
At $t=0$, $\bm{n}_{g}=(0,1,0)$ by construction, which lies on the $Y$-axis of the Bloch sphere.
Since the chiral-symmetry operator of $U_g(q,t)$ is $\sigma_y$, the time-evolved Bloch vector at any given quasimomentum $q$ tracks a meridian of the Bloch sphere, reaching $\bm{n}_{g}=(0,-1,0)$ at a momentum-dependent time $t_f(q)$. When the effective two-band model associated with gap $g$ is topologically nontrivial, the Bloch vector would cover the entire Bloch sphere as $q$ and $t$ vary across the $S^2$ manifold spanned by $q\in[0,2\pi)$ and $t\in (0,t_f(q)]$ [see Fig.~\ref{fig2}(b)]. 
The resulting $S^2\rightarrow S^2$ map is characterized by the gap-resolved Chern number
\begin{equation}
C_g=-
\frac{1}{4\pi}
\int_{\mathrm{BZ}} dq
\int_{t_0}^{t_{\mathrm f}(q)} dt\,
\bm n_g\cdot
\left(
\partial_q\bm n_g\times\partial_t\bm n_g
\right),
\label{eq:gap-skyrmion-number}
\end{equation}
which also corresponds to the skyrmion number of the emergent spin texture over the corresponding quasimomentum--time manifold. 
Note that the emergent skyrmions here are different from those in Ref.~\cite{yang2018dynamical}, as the underlying quasimomentum-time manifolds are chosen differently~\cite{chang2018topology,hsu2021disorder}.

We now show how the gap-resolved Chern numbers can be used to reconstruct the non-Abelian topology of the post-quench Hamiltonian.
As discussed above, the elementary charges $\pm k$ and $\pm i$ generate the quaternion group $Q_8$.
In the three-band system, a single skyrmion in gap 1 (gap 2) is characterized by $C_1=\pm 1$ ($C_2=\pm 1$), and carries the topological charge $\pm k$ ($\pm i$). 
More generally, the magnitude of the gap-resolved Chern numbers $|C_1|$ and $|C_2|$ count the occurrences of the generators $\pm k$ and $\pm i$ (their signs consistent with those of $C_{1,2}$), respectively, in the decomposition of the total quaternion charge of the system. Their distribution over the gaps therefore correctly predict the edge-mode patterns, when the system is under the open boundary condition. 
On the other hand, the product of these gap-resolved generators should reproduce the total quaternion charge except for the overall sign, since the order of their product is unspecified.
But this overall sign can be determined from $(C_1,C_2)$, as tabulated in Table~\ref{tab:q8-skyrmion}. This is because 
$(C_1,C_2)$ fully characterizes the eigenframe rotations, which encode  
complete information of the non-Abelian topology~\cite{supplemental}.

As concrete examples, from the gap-resolved spin textures
$\bm n_g(q,t)$ shown in Fig.~\ref{fig2}(c),
we obtain $(C_1,C_2)=(0,1)$, $(1,0)$, $(1,1)$, and $(2,0)$, 
which are characterized by the total quaternion charges $i$, $k$, $j$, and $-1$, respectively, 
according to Tab.~I.
In particular, for the case with charge $j$, the decomposition is necessarily $j=ki$, where the generators $k$ and $i$, as well as the positive sign of the total charge, are all given by $(C_1,C_2)=(1,1)$.
Likewise, the decomposition with charge $-1$ is given by $-1=k^2$, for $(C_1,C_2)=(2,0)$.
These decompositions also agree with the corresponding edge-mode configurations
in Fig.~\ref{fig1}(b).

{\it Dynamic signatures of topological transitions.---}
The emergent spin textures characterized by gap-resolved Chern numbers provide dynamic signatures for the global non-Abelian topology. For an illustration, we show the evolution of the gap-resolved spin textures, as the parameters of the post-quench Hamiltonian are tuned across topological phase transitions. We consider a smooth interpolation 
\begin{equation}
  H_\lambda(q)=(1-\lambda)H_{k}(q)+\lambda H_{i}(q),
  \label{eq:k-to-i-path}
\end{equation}
where $\lambda\in[0,1]$ continuously connects Hamiltonians $H_{k}$ and $H_{i}$, characterized by total quaternion charges $k$ and $i$, respectively. The detailed models and their parameters are given in the Supplemental Material~\cite{supplemental}.

In Fig.~\ref{fig3}(a), we plot the open-boundary spectrum as a function of $\lambda$, which reveals two successive topological transitions. For $\lambda<\lambda_c^{(1)}=1/3$, the system is still characterized by $k$, with a topologically nontrivial gap 1 hosting a pair of edge modes (shown in red). 
At $\lambda_c^{(1)}$, gap 1 closes and subsequently reopens into a topologically trivial phase, where the topological edge modes are absent. 
Gap 2 then closes and reopens at $\lambda_c^{(2)}=2/3$, beyond which topological edge modes emerge in gap 2, signaling a transition into the phase characterized by charge $i$. 

The five columns in Fig.~\ref{fig3}(b) show the gap-resolved
spin textures $\bm n_{1,2}(q,t)$ at five representative values of $\lambda$.
For each $\lambda$, we construct the full three-band evolution operator
$U_\lambda(q,t)=\exp[-\ii H_\lambda(q)t]$ and extract $\bm n_1(q,t)$ and
$\bm n_2(q,t)$ using Eq.~\eqref{eq:gap-bloch-vector}. The upper and lower rows
therefore show the spin textures associated with gap 2 and gap 1,
respectively, of the same quench dynamics.
At the critical points $\lambda_c^{(1)}$ and $\lambda_c^{(2)}$, $t_\mathrm{f}$ diverges at the gap-closing point $q=0$, and the Chern numbers become ill-defined. Otherwise, $(C_1,C_2)$ change from $(1,0)$ to $(0,0)$ and eventually to $(0,1)$, across the phase transition points. These signatures thus provide
a dynamical probe for topological phase transitions.

{\it Discussion.---}
We have developed a gap-resolved quench scheme for probing non-Abelian topology in one-dimensional three-band systems. 
Focusing on the contribution from each energy gap between adjacent bands, we show that gap-resolved Chern numbers over the emergent quasimomentum--time manifold directly correspond to the edge-mode patterns and enable the reconstruction of the global non-Abelian topological charge. 
These emergent Chern numbers manifest as dynamic skyrmions in the projected spin textures, thus offering a practical dynamic detection scheme for the underlying non-Abelian topology.

While we focus on three-band models in this study, our idea can be extended to general multiband non-Abelian topological insulators~\cite{wu2019nonabelian,jiang2021fourband,jiang2026sixband}. Therein, knowledge of the gap-resolved Chern numbers alone is often insufficient to fully capture the eigenframe rotations in higher-dimensional spaces, for which additional measurements are needed.
In four-band non-Abelian topological insulators~\cite{jiang2021fourband}, for instance, the global non-Abelian topology can still be recovered from the gap-resolved Chern numbers with the aid of additional quench processes~\cite{supplemental}.
Our approach can also be generalized to non-Hermitian settings~\cite{ding2022nh,supplemental}, 
where the PT symmetry, now ensuring the realness of both the eigenspectrum and eigenvectors, can still protect the non-Abelian topology. Given the recent experimental progress in non-Abelian topological insulators~\cite{guo2021charges,jiang2021fourband,jiang2026sixband,lin2026floquet,qiu2026floquet}, we expect our scheme should be readily accessible to synthetic physical systems such as acoustic materials~\cite{xue2022acoustics}, photons~\cite{ozawa2019photonics}, and cold atoms~\cite{cooper2019ultracold}.

\begin{acknowledgments}
This work is supported by the National Natural Science Foundation of China (Grant Nos. 12374479, 12504189), and Quantum Science and Technology-National Science and Technology Major Project (Grant No. 2021ZD0301904). T.L. is supported by the Guangdong Basic and Applied Basic Research Foundation (Grant No.2026A1515012098).
\end{acknowledgments}

\clearpage
\onecolumngrid
\setcounter{page}{1}
\renewcommand{\thepage}{S\arabic{page}}
\providecommand{\toprule}{\hline}
\providecommand{\midrule}{\hline}
\providecommand{\bottomrule}{\hline}

\setcounter{section}{0}
\setcounter{subsection}{0}
\setcounter{equation}{0}
\setcounter{figure}{0}
\setcounter{table}{0}
\setcounter{secnumdepth}{2}
\renewcommand{\thesection}{S\arabic{section}}
\renewcommand{\thesubsection}{S\arabic{section}.\arabic{subsection}}
\makeatletter
\renewcommand{\p@subsection}{}
\makeatother
\renewcommand{\theequation}{S\arabic{equation}}
\renewcommand{\thefigure}{S\arabic{figure}}
\renewcommand{\thetable}{S\arabic{table}}
\renewcommand{\theHsection}{S\arabic{section}}
\renewcommand{\theHsubsection}{S\arabic{section}.\arabic{subsection}}
\renewcommand{\theHequation}{S\arabic{equation}}
\renewcommand{\theHfigure}{S\arabic{figure}}
\renewcommand{\theHtable}{S\arabic{table}}

\clearpage
\onecolumngrid
\suppressfloats[t]
\begin{center}
{\large\bfseries Supplemental Material for ``Probing Non-Abelian Topology through Quench Dynamics''}
\end{center}

In this Supplemental Material, we provide details of the tight-binding Hamiltonians, the model parameters, and calculations of the gap-resolved Chern numbers and the construction of global topological charge. We also discuss extensions to non-Hermitian and four-band systems.

\section{Three-band tight-binding model and parameters}
\label{sm:sec:three-band}

We use the sublattice order $(A,B,C)$. Fourier transforming the tight-binding
Hamiltonian in Eq.~(1) of the main text, including
nearest-neighbor (NN) and next-nearest-neighbor (NNN) terms, gives the real and
symmetric Bloch Hamiltonian
\begin{equation}
H(q)=
\begin{pmatrix}
d_A(q)&h_{AB}(q)&h_{AC}(q)\\
h_{AB}(q)&d_B(q)&h_{BC}(q)\\
h_{AC}(q)&h_{BC}(q)&d_C(q)
\end{pmatrix},
\label{sm:eq:three-band-hamiltonian}
\end{equation}
where
\begin{align}
d_X(q)&=s_{XX}+2v_{XX}\cos q+2v'_{XX}\cos 2q,
\qquad X=A,B,C,\nonumber\\
h_{AB}(q)&=2u\sin q+2u'\sin 2q,\nonumber\\
h_{BC}(q)&=2v\sin q+2v'\sin 2q,\nonumber\\
h_{AC}(q)&=2w\sin q+2w'\sin 2q.
\label{sm:eq:three-band-elements}
\end{align}
The sine form follows from purely imaginary directed hopping amplitudes in
real space and makes $H(q)$ real in the chosen PT-symmetric representation.
The model parameters are taken from Ref.~\cite{guo2021charges}.
Parameters not displayed in Tables~\ref{sm:tab:three-band-nn} and
\ref{sm:tab:three-band-nnn} are zero.

\begin{table}
\caption{Ideal NN three-band flat-band parameters. All NNN coefficients vanish.}
\label{sm:tab:three-band-nn}
\centering
\renewcommand{\arraystretch}{1.15}
\setlength{\tabcolsep}{5.5pt}
\begin{tabular}{c c c c c c c c c c}
\toprule
charge & $s_{AA}$ & $s_{BB}$ & $s_{CC}$ & $v_{AA}$ & $v_{BB}$ & $v_{CC}$ & $u$ & $v$ & $w$\\
\midrule
$+i$ & $1$ & $5/2$ & $5/2$ & $0$ & $1/4$ & $-1/4$ & $0$ & $1/4$ & $0$\\
$-i$ & $1$ & $5/2$ & $5/2$ & $0$ & $1/4$ & $-1/4$ & $0$ & $-1/4$ & $0$\\
$+j$ & $2$ & $2$ & $2$ & $1/2$ & $0$ & $-1/2$ & $0$ & $0$ & $-1/2$\\
$-j$ & $2$ & $2$ & $2$ & $1/2$ & $0$ & $-1/2$ & $0$ & $0$ & $1/2$\\
$+k$ & $3/2$ & $3/2$ & $3$ & $1/4$ & $-1/4$ & $0$ & $1/4$ & $0$ & $0$\\
$-k$ & $3/2$ & $3/2$ & $3$ & $1/4$ & $-1/4$ & $0$ & $-1/4$ & $0$ & $0$\\
\bottomrule
\end{tabular}
\end{table}

\begin{table}
\caption{Ideal NNN representative of $-1=k^2$.  All NN coefficients vanish.}
\label{sm:tab:three-band-nnn}
\centering
\renewcommand{\arraystretch}{1.15}
\setlength{\tabcolsep}{5.5pt}
\begin{tabular}{c c c c c c c c c c}
\toprule
charge & $s_{AA}$ & $s_{BB}$ & $s_{CC}$ & $v'_{AA}$ & $v'_{BB}$ & $v'_{CC}$ & $u'$ & $v'$ & $w'$\\
\midrule
$-1=k^2$ & $3/2$ & $3/2$ & $3$ & $-1/4$ & $1/4$ & $0$ & $-1/4$ & $0$ & $0$\\
\bottomrule
\end{tabular}
\end{table}

The interpolation used in Fig.3 of the main text is obtained by linearly
interpolating every coefficient between the $k$ and $i$ rows of
Table~\ref{sm:tab:three-band-nn}.  Its nonzero parameters are
\begin{align}
s_{AA}&=\frac32-\frac{\lambda}{2},&
s_{BB}&=\frac32+\lambda,&
s_{CC}&=3-\frac{\lambda}{2},\nonumber\\
v_{AA}&=\frac{1-\lambda}{4},&
v_{BB}&=-\frac14+\frac{\lambda}{2},&
v_{CC}&=-\frac{\lambda}{4},\nonumber\\
u&=\frac{1-\lambda}{4},&
v&=\frac{\lambda}{4},&
w&=0.
\label{sm:eq:k-i-parameters}
\end{align}
The two gap closings occur at $q=0$: the lower gap closes at
$\lambda=1/3$ and the upper gap at $\lambda=2/3$, as marked in Fig.3 of the main text.

\section{Correspondence between gap-resolved Chern numbers and quaternion charges}
\label{sm:sec:q8-skyrmion-correspondence}

For any topological space $X$, the homotopy groups $\pi_n(X)$ are
Abelian for $n\geq2$.
Consequently, the Chern number of a single quasimomentum--time map
is an Abelian invariant and does not directly define a non-Abelian
charge. We instead use the gap-resolved pair
$(C_1,C_2)\in\mathbb Z^2$ to distinguish all eight elements of $Q_8$
for the models in Table~I of the main text.
The reason is that $C_g$ records the eigenframe rotation: its sign
indicates the direction, and its magnitude characterizes the amount
of rotation. Moreover, $C_1$ and $C_2$
are related because the eigenvectors must remain orthogonal.

\begin{figure}
\centering
\includegraphics[width=0.98\textwidth]{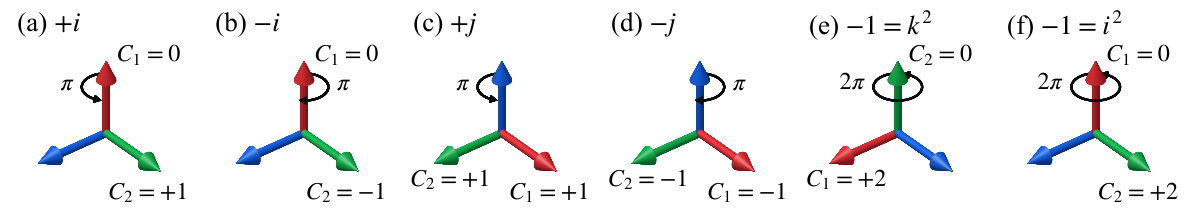}
\caption{Eigenframe rotations and gap-resolved Chern numbers.
Red, blue, and green arrows denote $u_1$, $u_2$, and $u_3$, respectively,
as in Fig.~1(a) of the main text. Black curved arrows indicate the rotation direction as quasimomentum traverses the Brillouin zone, with the rotation angle labelled alongside. $C_1$ and $C_2$ are marked beside $u_1$ and $u_3$,
respectively.
(a,b) Rotations by $\pi$ in opposite directions about $u_1$ give
$\pm i$, with $C_1=0$ and $C_2=\pm1$.
(c,d) Rotations by $\pi$ in opposite directions about $u_2$ give
$\pm j$, with $C_1=C_2=\pm1$.
(e,f) Rotations by $2\pi$ about $u_3$ and $u_1$ give $-1=k^2$ and
$-1=i^2$, with $(C_1,C_2)=(2,0)$ and $(0,2)$, respectively.}
\label{sm:fig:eigenframe-rotations}
\end{figure}

For example, for $\pm i$, $u_2$ and $u_3$ rotate by $\pi$ about $u_1$
[Fig.~\ref{sm:fig:eigenframe-rotations}(a,b)]. Hence $C_1=0$, whereas
$C_2=+1$ or $-1$ distinguishes the two rotation directions under our
sign convention. The $\pm k$ cases are analogous, with the eigenframe
rotating about $u_3$, giving $C_1=\pm1$ and $C_2=0$.
For $\pm j$, $u_1$ and $u_3$ rotate about $u_2$
[Fig.~\ref{sm:fig:eigenframe-rotations}(c,d)]. Since $u_1$ and
$u_3$ must remain orthogonal, they must rotate together: reversing
one while keeping the motion of the other would violate orthogonality.
Reversing the frame rotation therefore reverses both signs, giving
$(C_1,C_2)=(1,1)$ for $+j$ and $(-1,-1)$ for $-j$.
The magnitude also carries rotation information. A $2\pi$ rotation
about $u_3$ gives $(C_1,C_2)=(2,0)$, while a $2\pi$ rotation about
$u_1$ gives $(0,2)$ [Fig.~\ref{sm:fig:eigenframe-rotations}(e,f)].
Both have quaternion charge $-1$, but the two pairs distinguish
which gap is involved. Together, these examples explain the
rotation information contained in $C_g$ and the relation between
$C_1$ and $C_2$ imposed by orthogonality. The calculation connecting
these rotations to the Chern numbers is given in
Sec.~\ref{sm:sec:gap-proof}.

\section{Calculation of topological invariants}
\subsection{Zak phase}

For an isolated Bloch band $n$, the Zak phase is the Berry phase accumulated as the quasimomentum traverses the Brillouin zone~\cite{zak1989berry}. It is defined as
\begin{equation}
\phi_n
=
i\oint_{\mathrm{BZ}} dq\,
\langle u_n(q)|\partial_q u_n(q)\rangle
\quad (\mathrm{mod}\ 2\pi),
\label{eq:Zak_band}
\end{equation}
Here $|u_n(q)\rangle$ denotes the $n$th Bloch eigenstate and $\langle u_n(q)|$ its Hermitian conjugate.  For a PT-symmetric Hamiltonian that admits a real representation, the nontrivial value $\phi_n=\pi$ corresponds to a sign reversal of the real eigenstate after parallel transport around the Brillouin zone.

For a given band gap, the gap Zak phase is defined as the sum of the Zak phases of all bands below that gap,
\begin{equation}
\phi_{\mathrm{gap},m}
=
\sum_{n=1}^{m}\phi_n
\quad (\mathrm{mod}\ 2\pi),
\label{eq:Zak_gap}
\end{equation}
where the $m$th gap separates bands $m$ and $m+1$. For a three-band system, the two
gap Zak phases are therefore
\begin{equation}
\phi_{\mathrm{gap},1}=\phi_1,
\qquad
\phi_{\mathrm{gap},2}=\phi_1+\phi_2
\quad (\mathrm{mod}\ 2\pi).
\end{equation}

\subsection{Non-Abelian topological charge}

\label{sm:sec:q8-charge}

For a three-band PT-symmetric Hamiltonian with open gaps, the eigenvectors
can be chosen real and arranged as an orthonormal frame
$F(q)=(|u_1(q)\rangle,|u_2(q)\rangle,|u_3(q)\rangle)$.  Independent sign changes of its columns are
gauge transformations, so a closed Brillouin-zone path is a loop in the real
flag manifold $O(3)/(\mathbb Z_2)^3$.  The Berry--Wilczek--Zee non-Abelian connection is
\cite{wu2019nonabelian,guo2021charges,zhang2026topoword,wilczek1984gauge}
\begin{equation}
\mathcal A_{mn}(q)=\langle u_m(q)|\partial_q u_n(q)\rangle,
\qquad \mathcal A(q)=F^T(q)\partial_qF(q).
\label{sm:eq:wbz}
\end{equation}
Differentiating $F^TF=I$ gives $\mathcal A^T=-\mathcal A$; hence
\begin{equation}
\mathcal A(q)\in\mathfrak{so}(3)
=\{X\in\mathbb R^{3\times3}:X^T=-X\}.
\label{sm:eq:so3-definition}
\end{equation}
Here $\mathfrak{so}(3)$ denotes the Lie algebra of $SO(3)$, equivalently its
tangent space at the identity.
We write
\begin{equation}
\mathcal A(q)=a_x(q)L_x+a_y(q)L_y+a_z(q)L_z,
\label{sm:eq:so3-components}
\end{equation}
where $L_x,L_y,L_z$ generate rotations in the $23$, $13$, and $12$ planes,
respectively.  The double cover $\mathrm{Spin}(3)\simeq SU(2)$ lifts these
generators according to
\begin{equation}
L_x\longmapsto \frac{\ii}{2}\sigma_x,\qquad
L_y\longmapsto \frac{\ii}{2}\sigma_y,\qquad
L_z\longmapsto \frac{\ii}{2}\sigma_z.
\label{sm:eq:so3-su2-lift}
\end{equation}
The lifted connection is
$\widetilde{\mathcal A}=\ii(a_x\sigma_x+a_y\sigma_y+a_z\sigma_z)/2$. The non-Abelian Wilson loop is
\begin{equation}
\widetilde W=\mathcal P\exp\left(\oint_{\rm BZ}
\widetilde{\mathcal A}(q)\,dq\right).
\label{sm:eq:lifted-wilson-loop}
\end{equation}
Here $\mathcal P$ is the momentum-ordering operator: matrices evaluated later
along the chosen orientation of the Brillouin-zone loop are placed to the
left.
Because the physical frame closes only up to independent column signs, the
allowed lifted holonomies are the eight matrices
\begin{equation}
\widetilde W\in
\{\,\pm I,\ \pm\ii\sigma_x,\ \pm\ii\sigma_y,\ \pm\ii\sigma_z\,\}.
\label{sm:eq:q8-matrices}
\end{equation}
They represent $Q_8$ through
\begin{equation}
\pm I\leftrightarrow\pm1,\qquad
\pm\ii\sigma_x\leftrightarrow\pm i,\qquad
\pm\ii\sigma_y\leftrightarrow\pm j,\qquad
\pm\ii\sigma_z\leftrightarrow\pm k.
\label{sm:eq:q8-map}
\end{equation}

\section{Mathematical structure of the gap-resolved reduction}
\label{sm:sec:gap-proof}

The main text defines the gap-dependent basis $W_g$ and the projected
evolution $U_g$ for both physical gaps.  Here we explain why this construction
retains a two-band topological invariant and why that invariant is measured by
the dynamical skyrmion structure.  Without loss of generality, we present the
derivation for gap 1.

\subsection{Effective chiral two-band dynamics}

In this section, we show that the normalized Bloch vector obtained from
the physical three-band evolution has an exact representation in terms
of chiral two-band dynamics.

For a PT-symmetric post-quench Hamiltonian, the physical evolution is
\begin{equation}
U(q,t)=e^{-\ii H(q)t}
=\sum_{\alpha=1}^{3}
e^{-\ii E_\alpha(q)t}
|u_\alpha(q)\rangle\langle u_\alpha(q)|.
\label{sm:eq:chiral-physical-u}
\end{equation}
We present the derivation for gap 1. The construction for gap 2 follows
by replacing the band-1 eigenstate and the complementary bands $(2,3)$
with the band-3 eigenstate and bands $(1,2)$, respectively.

We use the reference basis $|e_i\rangle=|u_i(q_0)\rangle$ fixed in the
main text. For gap 1, define
\begin{equation}
P_{23}=I_3-|e_1\rangle\langle e_1|,
\qquad
\rho(q)=\|P_{23}|u_1(q)\rangle\|.
\label{sm:eq:reference-plane}
\end{equation}
The two basis vectors are
\begin{equation}
|w_1(q)\rangle
=\chi_1(q)\frac{P_{23}|u_1(q)\rangle}{\rho(q)},
\qquad
W_1(q)=\bigl(|e_1\rangle,|w_1(q)\rangle\bigr).
\label{sm:eq:w1-recalled}
\end{equation}
Here $\chi_1(q)=\pm1$ is fixed by the orientation convention of the
complete eigenframe specified below.  At an isolated zero of $\rho$, the
second basis vector is defined by the continuous limit from the neighboring
momentum interval.
The real basis matrix $W_1(q)$ satisfies
$W_1^\dagger W_1=I_2$ and $W_1W_1^\dagger=P_{W_1}$, where $P_{W_1}$
projects onto the selected plane. Its columns span a plane containing
$|u_1(q)\rangle$. Thus,
\begin{equation}
|\psi_1(q)\rangle
=W_1^\dagger(q)|u_1(q)\rangle
=\begin{pmatrix}a(q)\\b(q)\end{pmatrix},
\qquad
a^2+b^2=1,
\qquad
|u_1\rangle=W_1|\psi_1\rangle.
\label{sm:eq:chiral-spinor}
\end{equation}
Here $a(q)=\langle e_1|u_1(q)\rangle$ and $b(q)=\chi_1(q)\rho(q)$
are real. Define the orthogonal state and its corresponding three-band
vector as
\begin{equation}
|\psi_{1\perp}\rangle
=\begin{pmatrix}-b\\a\end{pmatrix},
\qquad
|v_{23}\rangle=W_1|\psi_{1\perp}\rangle.
\label{sm:eq:chiral-complement}
\end{equation}
Since $\langle u_1|v_{23}\rangle=0$, the vector $|v_{23}\rangle$ lies
in the space spanned by $|u_2\rangle$ and $|u_3\rangle$. Below, the
dependence on $q$ is suppressed where no confusion arises.

The projected physical evolution is
\begin{equation}
U_1(q,t)=W_1^\dagger(q)U(q,t)W_1(q).
\label{sm:eq:chiral-projected-u}
\end{equation}
Because $|u_1\rangle$ is an eigenstate of $U$, its matrix elements
connecting $|u_1\rangle$ and $|v_{23}\rangle$ vanish. Therefore, with
$R_1=(|\psi_1\rangle,|\psi_{1\perp}\rangle)\in SO(2)$, we obtain
\begin{align}
R_1^T U_1 R_1
&=
\begin{pmatrix}
e^{-\ii E_1t}&0\\
0&z_{23}(q,t)
\end{pmatrix},
\label{sm:eq:chiral-diagonal-u}\\
z_{23}(q,t)
&=\langle v_{23}|U(q,t)|v_{23}\rangle
=\sum_{\alpha=2,3}
|\langle u_\alpha|v_{23}\rangle|^2
e^{-\ii E_\alpha t}.
\label{sm:eq:chiral-z23}
\end{align}
Equivalently,
\begin{equation}
U_1=e^{-\ii E_1t}P_1+z_{23}(I_2-P_1),
\qquad
P_1=|\psi_1\rangle\langle\psi_1|.
\label{sm:eq:chiral-projector-u}
\end{equation}
This expression follows directly from the physical evolution and does
not require $E_2=E_3$.

In general, $|z_{23}|\leq1$, and
\begin{equation}
U_1^\dagger U_1=P_1+|z_{23}|^2(I_2-P_1).
\label{sm:eq:chiral-projected-norm}
\end{equation}
Thus, $U_1$ need not be unitary: evolution can take part of
$|v_{23}\rangle$ outside the selected plane. The correspondence with
chiral two-band dynamics concerns the normalized Bloch vector
constructed from $U_1$.

To establish this correspondence, write
\begin{equation}
P_1=\frac{I_2+\bm m_1\cdot\bm\sigma}{2},
\qquad
\bm m_1
=\langle\psi_1|\bm\sigma|\psi_1\rangle
=(2ab,0,a^2-b^2).
\label{sm:eq:chiral-axis}
\end{equation}
The unit vector $\bm m_1(q)$ lies on the great circle defined by the
intersection of the Bloch sphere with the $XZ$ plane. Define
\begin{equation}
\mathcal N_1=U_1\sigma_yU_1^\dagger,
\qquad
n_{1,\mu}=\frac{1}{2}\operatorname{Tr}
(\mathcal N_1\sigma_\mu),
\qquad \mu=x,y,z.
\label{sm:eq:chiral-bloch-definition}
\end{equation}
Since $R_1^T\sigma_yR_1=\sigma_y$,
Eq.~\eqref{sm:eq:chiral-diagonal-u} gives
\begin{equation}
R_1^T\mathcal N_1R_1
=
\begin{pmatrix}
0&-\ii e^{-\ii E_1t}z_{23}^*\\
\ii e^{\ii E_1t}z_{23}&0
\end{pmatrix}.
\label{sm:eq:chiral-bloch-matrix}
\end{equation}
Consequently,
\begin{equation}
\operatorname{Tr}\mathcal N_1=0,
\qquad
\mathcal N_1^2=|z_{23}|^2I_2,
\qquad
|\bm n_1|=|z_{23}|.
\label{sm:eq:chiral-bloch-norm}
\end{equation}

On a time interval starting at $t=0$ where $z_{23}(q,t)$ remains
nonzero, define the continuously followed relative phase by
\begin{equation}
e^{-\ii E_1t}z_{23}^*(q,t)
=|z_{23}(q,t)|e^{-\ii\Phi_1(q,t)},
\qquad
\Phi_1(q,0)=0.
\label{sm:eq:chiral-relative-phase}
\end{equation}
The normalized Bloch vector then satisfies the exact identity
\begin{equation}
\frac{{\bm n}_1\cdot\bm\sigma}{|\bm n_1|}
=\frac{\mathcal N_1}{|z_{23}|}
=V_1\sigma_yV_1^\dagger,
\qquad
V_1(q,t)
=\exp\left[
-\frac{\ii\Phi_1(q,t)}{2}
\bm m_1(q)\cdot\bm\sigma
\right].
\label{sm:eq:chiral-normalized-dynamics}
\end{equation}
Thus, normalization removes the amplitude $|z_{23}|$, while retaining
the relative phase generated by the physical band energies.

The unitary matrix $V_1$ is precisely the evolution operator of the
two-band Hamiltonian
\begin{equation}
h_1^{\rm eff}(q,t)
=\frac{\Omega_1(q,t)}{2}
\left[
m_{1x}(q)\sigma_x+m_{1z}(q)\sigma_z
\right],
\qquad
\Phi_1(q,t)=\int_0^t\Omega_1(q,t')\,dt'.
\label{sm:eq:chiral-effective-generator}
\end{equation}
Here $\Omega_1(q,t)$ is the generally time-dependent precession rate,
while the axis $\bm m_1(q)$ is fixed at each momentum.
Because this Hamiltonian lies entirely in the $XZ$ plane, it satisfies
\begin{equation}
\{h_1^{\rm eff}(q,t),\sigma_y\}=0.
\label{sm:eq:chiral-symmetry}
\end{equation}
The normalized Bloch vector obtained from the physical three-band
evolution therefore coincides with that of a chiral two-band Hamiltonian with chiral operator $\sigma_y$.

More directly, the connection between the projected evolution and an effective chiral two-band Hamiltonian can be understood by spectrally
flattening the two sides of the selected gap while keeping the corresponding
eigenstates unchanged.  As long as gap 1 remains open, we may continuously deform the spectrum by assigning a constant energy $E_1$ to band 1 and a common reference energy $E_{23}>E_1$ to bands $(2,3)$, without closing gap 1. The resulting flattened Hamiltonian is
\begin{equation}
H_1^{\rm flat}(q)
=E_1|u_1(q)\rangle\langle u_1(q)|
+E_{23}\left[I-|u_1(q)\rangle\langle u_1(q)|\right].
\label{sm:eq:chiral-flat-hamiltonian}
\end{equation}
This deformation leaves the band-1 projector unchanged and therefore
preserves the gap-1 eigenstate topology used in the present construction.

In the flattened limit, the complementary amplitude becomes
$z_{23}=e^{-\ii E_{23}t}$, so that
$\Phi_1(q,t)=\Delta_1t$, where $\Delta_1=E_1-E_{23}$ is constant.
The selected plane is now invariant under the evolution, and
\begin{equation}
W_1^\dagger e^{-\ii H_1^{\rm flat}(q)t}W_1
=e^{-\ii\bar E_1t}
\exp\left[
-\frac{\ii\Delta_1t}{2}\bm m_1(q)\cdot\bm\sigma
\right],
\qquad
\bar E_1=\frac{E_1+E_{23}}{2}.
\label{sm:eq:chiral-flat-evolution}
\end{equation}
After removing the overall phase, this is precisely the evolution
generated by the time-independent two-band Hamiltonian
\begin{equation}
\widetilde h_1^{\rm flat}(q)
=\frac{\Delta_1}{2}
\left[m_{1x}(q)\sigma_x+m_{1z}(q)\sigma_z\right],
\qquad
\{\widetilde h_1^{\rm flat}(q),\sigma_y\}=0.
\label{sm:eq:chiral-flat-generator}
\end{equation}
Thus, after spectral flattening, the gap-1 construction is represented by
a time-independent chiral two-band Hamiltonian in the $XZ$ plane.

\subsection{Topology and the sign convention}

Because $|\psi_1\rangle$ is real and normalized, it can be written with one
angle,
\begin{equation}
a(q)=\cos\theta(q),\qquad b(q)=\sin\theta(q),\qquad
\bm m_1(q)=\bigl(\sin(2\theta(q)),0,\cos(2\theta(q))\bigr).
\label{sm:eq:quotient-angles}
\end{equation}
The effective Hamiltonian therefore traces the $x$--$z$ circle as $q$ runs
through the Brillouin zone.  The two spinors
$|\psi_1\rangle$ and $-|\psi_1\rangle$ represent the same physical
eigenstate, so $\theta$ and $\theta+\pi$ are identified.  The number of turns
made by $\bm m_1$ is
\begin{equation}
\nu_1=\frac{1}{2\pi}\oint_{\rm BZ}d(2\theta)
=\frac{\theta(q_0+2\pi)-\theta(q_0)}{\pi}\in\mathbb Z.
\label{sm:eq:framed-winding}
\end{equation}
This is the usual winding number of a real two-band Hamiltonian.  Within the
gap-resolved construction, it cannot change under a continuous deformation
for which the selected physical gap and the oriented eigenframe used to fix
the projection remain well defined.

The same angle $2\theta(q)$ also determines the Bloch-vector dynamics.
In the flattened limit, $E_{23}>E_1$, so $\Phi_1=(E_1-E_{23})t<0$
for $t>0$. With the increasing time coordinate
$\tau=-\Phi_1=(E_{23}-E_1)t$, Eq.~(23) of the Method
and the Bloch-vector normalization give
\begin{equation}
\bm n_1(q,\tau)=
\bigl(\cos(2\theta)\sin\tau,\ \cos\tau,\
-\sin(2\theta)\sin\tau\bigr),\qquad 0\leq\tau\leq\pi .
\label{sm:eq:suspended-bloch-vector}
\end{equation}
All momenta start at $+\hat{\bm y}$ and end at $-\hat{\bm y}$, so the two
time boundaries each represent one point on the Bloch sphere.  Between them,
$2\theta(q)$ sets the azimuthal angle of the Bloch vector.  Explicitly,
\begin{equation}
\bm n_1\cdot(\partial_q\bm n_1\times\partial_\tau\bm n_1)
=-\sin\tau\,\partial_q(2\theta),
\label{sm:eq:skyrmion-density-proof}
\end{equation}
and the Chern number is therefore
\begin{equation}
C_1=-\frac{1}{4\pi}\int_{\rm BZ}dq\int_0^\pi d\tau\,
\bm n_1\cdot(\partial_q\bm n_1\times\partial_\tau\bm n_1)
=\frac{1}{2\pi}\oint_{\rm BZ}d(2\theta)=\nu_1 .
\label{sm:eq:skyrmion-winding-equivalence}
\end{equation}
Thus the Chern number of the gap-resolved quench is the winding number of
the effective real two-band Hamiltonian, with the overall sign fixed by the
momentum, time, and $\sigma_y$ conventions, including the minus sign
in Eq.~(6) of the main text.

It is useful to see geometrically what information is contained in this
integer.  If one ignores the sign $\chi_1$ and simply uses $b=\rho\geq0$,
then the effective state only records whether $|u_1(q)\rangle$ crosses the reference
$23$ plane:
\begin{equation}
a(q)=0\ \Longleftrightarrow\ |u_1(q)\rangle\in\Pi_{23},
\label{sm:eq:plane-crossing}
\end{equation}
where $\Pi_{23}=\operatorname{span}\{|e_2\rangle,|e_3\rangle\}$.
Counting such crossings without assigning a direction gives only their
parity,
\begin{equation}
N_1\bmod2=w_1[\mathcal L_1]\in\mathbb Z_2,
\qquad \mathcal L_1(q)=\operatorname{span}\{|u_1(q)\rangle\},
\label{sm:eq:crossing-parity}
\end{equation}
which is the familiar first Stiefel--Whitney number of the real band-1
eigenline~\cite{ahn2019stiefel}.  In simple terms, it distinguishes an even number of crossings
from an odd number, but it cannot distinguish two crossings made in the same
direction from two crossings that cancel.

The norm $\rho$ records how far the eigenvector moves into the projected
plane but does not retain the orientation inherited from the complete
eigenframe.  The sign $\chi_g$ supplies this information.  We use the
eigenvectors at the reference momentum, $|e_n\rangle=|u_n(q_0)\rangle$, as
fixed coordinate axes and determine the sign from the oriented motion of
$|u_1(q)\rangle$ or $|u_3(q)\rangle$ relative to these axes.

For gap 1, write
\begin{equation}
 P_{23}|u_1(q)\rangle
 =c_2(q)|e_2\rangle+c_3(q)|e_3\rangle,
 \qquad c_j(q)=\langle e_j|u_1(q)\rangle.
\label{sm:eq:chi-gap1-components}
\end{equation}
For the representative trajectories used in this work, the orientation
convention assigns a positive sign to motion toward $+|e_2\rangle$ or
$-|e_3\rangle$.  This gives
\begin{equation}
 \chi_1(q)=
 \begin{cases}
  \operatorname{sgn}c_2(q),&|c_2(q)|>|c_3(q)|,\\
  -\operatorname{sgn}c_3(q),&|c_3(q)|>|c_2(q)|.
 \end{cases}
\label{sm:eq:chi-gap1-rule}
\end{equation}
At isolated points where the projected vector vanishes,  $b=\chi_1\rho$ and the basis vector $|w_1\rangle$ are defined by
continuation from the neighboring momentum interval.

For gap 2, write
\begin{equation}
 P_{12}|u_3(q)\rangle
 =d_1(q)|e_1\rangle+d_2(q)|e_2\rangle,
 \qquad d_j(q)=\langle e_j|u_3(q)\rangle.
\label{sm:eq:chi-gap2-components}
\end{equation}
For gap 2, the same orientation convention assigns a positive sign to motion
toward $+|e_1\rangle$ or $-|e_2\rangle$, giving
\begin{equation}
 \chi_2(q)=
 \begin{cases}
  \operatorname{sgn}d_1(q),&|d_1(q)|>|d_2(q)|,\\
  -\operatorname{sgn}d_2(q),&|d_2(q)|>|d_1(q)|.
 \end{cases}
\label{sm:eq:chi-gap2-rule}
\end{equation}

The integer and parity descriptions are related by
\begin{equation}
\nu_1\bmod2=w_1[\mathcal L_1].
\label{sm:eq:z-to-z2}
\end{equation}
An elementary $k$ loop has $|\nu_1|=1$, while the double winding
$k^2=-1$ has $|\nu_1|=2$.  The latter is even and is therefore invisible to
the $\mathbb Z_2$ parity alone, but it is retained by the coordinate
$b$.  In this sense $\nu_1$ records repeated letters in the
selected gap, while the ordered information from all gaps is still needed to
recover the full non-Abelian charge.

\begin{figure}
\centering
\includegraphics[width=0.98\textwidth]{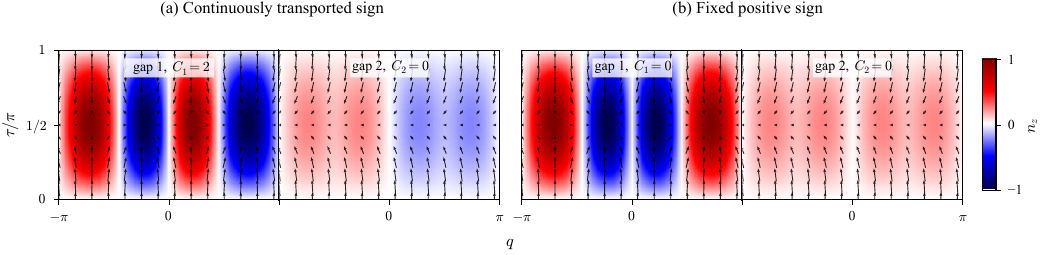}
\caption{Effect of the sign choice for the case $k^2=-1$.
(a) The oriented-frame convention preserves two windings with the same
orientation in gap 1 and gives $C_1=2$, while gap 2 remains trivial.
(b) Fixing the projected coefficient to be positive replaces the
coordinate $b$ by $|b|$.  The second half of the lower-gap Bloch-vector trajectory then
retraces the first in the opposite direction, giving $C_1=0$; gap 2 is again
trivial.  Color represents $n_z$, and arrows show $(n_x,n_y)$.}
\label{sm:fig:k2-sign-comparison}
\end{figure}

Figure~\ref{sm:fig:k2-sign-comparison} illustrates this distinction for
the double winding $k^2=-1$. With $\chi_1$ fixed by the oriented-frame
convention, the gap-1 Bloch-vector trajectory completes two turns with the
same orientation, giving $C_1=2$, in
agreement with Eqs.~\eqref{sm:eq:framed-winding} and
\eqref{sm:eq:skyrmion-winding-equivalence}.  If one instead fixes
$\chi_1=+1$, so that $b=\rho\geq0$, the second half of the Bloch-vector trajectory retraces
the first, so the contributions to the integral cancel, giving $C_1=0$.  Gap 2 is trivial in
both constructions.  The comparison therefore shows directly which
information is lost when the eigenframe orientation is discarded.

\section{Non-Hermitian extension and its limitations}
\label{sm:sec:non-hermitian}

We use the ideal $k$ model to explore how the gap-resolved picture changes
when the Hamiltonian becomes non-Hermitian.  An intracell nonreciprocal
coupling is added on the $BC$ link,
\begin{equation}
H_{\rm NH}=H_k
+\gamma\sum_n\left(
c^\dagger_{B,n}c_{C,n}-c^\dagger_{C,n}c_{B,n}
\right).
\label{sm:eq:nh-hamiltonian}
\end{equation}
Here $H_k$ is the tight-binding Hamiltonian in
Eq.~(1) of the main text with the ideal
$k$-phase parameters listed in Table~\ref{sm:tab:three-band-nn}.
The index $n$ labels the unit cells, and $\gamma$ is real.
It changes the two intracell hopping amplitudes in opposite directions while
preserving PT symmetry.  The Hamiltonian is nevertheless non-Hermitian for
$\gamma\ne0$~\cite{ding2022nh}.

In the PT-unbroken regime, the quaternion-group classification remains well
defined in the biorthogonal description, and the gap-resolved construction of
the main text can be applied using the corresponding left and right
eigenstates.

\begin{figure}
\centering
\includegraphics[width=0.99\textwidth]{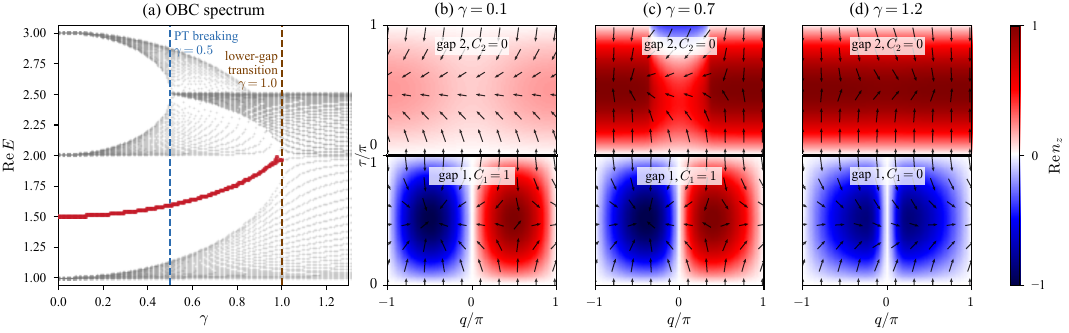}
\caption{Non-Hermitian spectrum and gap-resolved spin textures from the flattened dynamics of the
$+k$ model with the $BC$ perturbation in
Eq.~\eqref{sm:eq:nh-hamiltonian}.  (a) Real part of the OBC spectrum for a
24-cell chain as a function of $\gamma$.  Gray points are the OBC levels and
red points mark the two lower-gap boundary modes.  Dashed lines indicate PT breaking at
$\gamma_{\rm PT}=0.5$ and the lower-gap transition at $\gamma_c=1$.
(b)--(d) Gap-resolved spin textures at
$\gamma=0.1$, $0.7$, and $1.2$, respectively.
The vertical axis denotes $\tau=\Delta t$, where $\Delta$ is the energy gap of the flattened Hamiltonian in Eq.~\eqref{sm:eq:nh-flattened-hamiltonian}.
Color shows
$\operatorname{Re}n_z$ and arrows show the normalized direction of
$(\operatorname{Re}n_x,\operatorname{Re}n_y)$.
The value of $C_i$ is the Chern number calculated from the quench dynamics generated by the flattened gap-resolved Hamiltonian for gap $i$ according to Eq.~\eqref{sm:eq:nh-flattened-hamiltonian}.}
\label{sm:fig:nh-flattened}
\end{figure}

Figure~\ref{sm:fig:nh-flattened}(a) shows the real part of the
open-boundary spectrum.  For $0\leq\gamma<0.5$, PT symmetry is unbroken and
the spectrum is real.  The red levels are the two nearly degenerate boundary
modes in the lower gap.  At $\gamma_{\rm PT}=0.5$, the upper two bulk bands
meet at exceptional points at $q=\pm\pi$ and subsequently acquire complex-conjugate
energies over part of the Brillouin zone.  The eigenstates can then no longer
be organized into the real three-band frame required for a quaternion charge,
so a global $Q_8$ label is no longer well defined.  This does not immediately remove the boundary modes: the spectral
separation between the lowest band and the upper two bands remains open, and
the red branches persist throughout this part of the PT-broken regime.  At
$\gamma_c=1$, this lower spectral gap closes at $q=0$.  The boundary modes
merge into the bulk at this second transition and are absent for
$\gamma>1$.

To compare this spectral evolution with the gap-resolved quench dynamics, we first flatten
the energies on the two sides of the selected gap while keeping their right
and left eigenprojectors.  This removes the exponential amplification and
decay caused by $\operatorname{Im}E$ but preserves the eigenstate structure
that changes at the topological transition. The two flattened Hamiltonians are
\begin{align}
H_{\mathrm{gap}\,1}^{\mathrm{flat}}(q)
&=E_1|u_1^R(q)\rangle\langle u_1^L(q)|
 +E_{23}\left[I-|u_1^R(q)\rangle\langle u_1^L(q)|\right],\nonumber\\
H_{\mathrm{gap}\,2}^{\mathrm{flat}}(q)
&=E_3|u_3^R(q)\rangle\langle u_3^L(q)|
 +E_{12}\left[I-|u_3^R(q)\rangle\langle u_3^L(q)|\right].
\label{sm:eq:nh-flattened-hamiltonian}
\end{align}
Here $\langle u_n^L|u_m^R\rangle=\delta_{nm}$.  For the gap-resolved spin textures shown in
Fig.~\ref{sm:fig:nh-flattened}, we take $E_1=E_{12}=-1$ and
$E_{23}=E_3=+1$, so that the bands below and above the selected gap have
energies $-1$ and $+1$, respectively.  The flattening changes only the
eigenvalues; the biorthogonal projectors are unchanged, so the eigenstate
geometry entering the gap-resolved construction is preserved.

The resulting $2\times2$ evolution is used to construct the biorthogonal
Bloch vector: the state is evolved from the right and its dual partner from
the left~\cite{wang2019skyrmions,qiu2019fixed}.  In the flattened two-level problem, this may be written compactly as
\begin{equation}
n_{g,\mu}(q,\tau)=\frac{1}{2}\operatorname{Tr}
\left[U_g^R(q,\tau)\sigma_yU_g^L(q,\tau)\sigma_\mu\right],
\qquad U_g^L=(U_g^R)^{-1}.
\label{sm:eq:nh-biorthogonal-bloch-vector}
\end{equation}
The left--right normalization removes the arbitrary magnitude assigned to a
non-Hermitian eigenstate.  It should be distinguished from evolving with the
original complex energies, which we discuss below.

The spin textures in panels (b)(c)(d) follow the lower-gap physics in panel (a).
Before PT breaking, Fig.~\ref{sm:fig:nh-flattened}(b) reproduces the
Hermitian pattern $C_1=1$ in gap 1 and $C_2=0$ in gap 2.  At $\gamma=0.7$, the
spectrum is already partly complex, but the lower gap and its boundary modes
remain intact; accordingly, panel (c) retains the gap-1 response with $C_1=1$.
The flattened gap-resolved Bloch-vector construction therefore distinguishes
PT breaking from the gap-1 topological transition.  After the lower gap
closes, panel (d) instead gives $C_1=0$ in gap 1 at $\gamma=1.2$, consistently
with the disappearance of the red boundary branches.  The comparison shows
that the flattened construction can continue to track the gap-1 structure
encoded by the biorthogonal projectors inside a PT-broken spectrum.

The distinction between the flattened construction in Eq.~\eqref{sm:eq:nh-flattened-hamiltonian}
and the direct quench dynamics generated by $H_{\rm NH}$ becomes important
upon PT-symmetry breaking.
In the PT-broken regime, the eigenenergies acquire imaginary parts,
leading to exponential amplification and suppression of different components
during time evolution.
Consequently, the Chern numbers obtained from the direct quench dynamics
are no longer guaranteed to remain quantized.
By contrast, in the PT-unbroken regime, the energy spectrum remains real and no exceptional points arise.
The gap-resolved quench protocol therefore remains applicable directly
to $H_{\rm NH}$, yielding quantized Chern numbers without spectral flattening.

\section{Four-band extension}
\label{sm:sec:four-band}

\subsection{Four-band Hamiltonian and parameter sets}
\label{sm:sec:four-band-parameters}

We consider a one-dimensional PT-symmetric four-band model.
Following Ref.~\cite{jiang2021fourband}, we write the four-band real-space
tight-binding Hamiltonian with intracell, nearest-neighbor (NN), and
next-nearest-neighbor (NNN) terms as
\begin{equation}
\begin{aligned}
H_4={}&\sum_{n,X,Y}s_{XY}c_{X,n}^\dagger c_{Y,n}\\
&+\sum_{n,X,Y}\bigl(v_{XY}c_{X,n}^\dagger c_{Y,n+1}
+v'_{XY}c_{X,n}^\dagger c_{Y,n+2}+\mathrm{H.c.}\bigr),
\end{aligned}
\label{sm:eq:four-band-real-space}
\end{equation}
where $X,Y\in\{A,B,C,D\}$. The real symmetric coefficients
$s_{XY}=s_{YX}$ describe intracell terms, with $s_{XX}$ the onsite energies.
The diagonal hoppings $v_{XX}$ and $v'_{XX}$ are real.
For $X\neq Y$, we write the inter-sublattice hopping coefficients as
\begin{align}
v_{XY}=v_{YX}&=r_{XY}+\ii w_{XY},\\
v'_{XY}=v'_{YX}&=r'_{XY}+\ii w'_{XY},
\end{align}
with real $r_{XY},w_{XY},r'_{XY},w'_{XY}$. Unprimed and primed hopping
coefficients connect cells separated by one and two lattice constants,
respectively.

After Fourier transformation, the Bloch Hamiltonian in quasimomentum space is
the real symmetric matrix
\begin{equation}
H(q)=
\begin{pmatrix}
H_{AA}(q)&H_{AB}(q)&H_{AC}(q)&H_{AD}(q)\\
H_{AB}(q)&H_{BB}(q)&H_{BC}(q)&H_{BD}(q)\\
H_{AC}(q)&H_{BC}(q)&H_{CC}(q)&H_{CD}(q)\\
H_{AD}(q)&H_{BD}(q)&H_{CD}(q)&H_{DD}(q)
\end{pmatrix}.
\label{sm:eq:four-band-bloch-matrix}
\end{equation}

The diagonal and off-diagonal matrix elements are
\begin{align}
H_{XX}(q)&=s_{XX}+2v_{XX}\cos q+2v'_{XX}\cos2q,
\label{sm:eq:four-band-diagonal}\\
H_{XY}(q)&=s_{XY}+2r_{XY}\cos q+2r'_{XY}\cos2q
+2w_{XY}\sin q+2w'_{XY}\sin2q,
\quad X\neq Y.
\label{sm:eq:four-band-offdiagonal}
\end{align}

When all three gaps are open, the real symmetric Hamiltonian has four
orthonormal real eigenstates, which we order by increasing eigenenergy
to form the eigenframe $F(q)=[u_1(q),u_2(q),u_3(q),u_4(q)]\in O(4)$.
Each eigenstate is defined up to a sign, so the physical eigenframe
space is $O(4)/(\pm1)^4$. As $q$ traverses the Brillouin zone, the
eigenframe traces a loop in this space. Its topological charge belongs
to the fundamental group, the 16-element non-Abelian
group~\cite{jiang2021fourband}

\begin{equation}
Q_{16}=\{\pm1,\pm q_{12},\pm q_{13},\pm q_{14},
\pm q_{23},\pm q_{24},\pm q_{34},\pm q_{1234}\}.
\label{sm:eq:q16-elements}
\end{equation}

Here $\pm q_{ij}$ describe opposite $\pi$ rotations of the eigenframe
in the plane associated with bands $i$ and $j$. The charge $+1$ is
trivial, while $-1=q_{ij}^2$ corresponds to a $2\pi$ rotation.
The charges $\pm q_{1234}$ can be realized by rotations in two
orthogonal planes. Their different decompositions and the multiplication
rules of $Q_{16}$ are given in Sec.~\ref{sm:sec:q16}.

Tables~\ref{sm:tab:four-band-ideal} and
\ref{sm:tab:four-band-decompositions} give the ideal flat-band parameters
from Ref.~\cite{jiang2021fourband} for these topological charges.
In these tables, only the diagonal onsite coefficients $s_{XX}$,
diagonal NN cosine coefficients $v_{XX}$, and selected NN sine
coefficients $w_{XY}$ are nonzero. An unlisted $w_{XY}$ is zero.

\begin{table}
\caption{Ideal NN four-band flat-band parameters for the elementary
$q_{ij}$ charges and the trivial $+1$ model.}
\label{sm:tab:four-band-ideal}
\centering
\begingroup
\normalsize
\setlength{\tabcolsep}{3.2pt}
\renewcommand{\arraystretch}{1.10}
\begin{tabular}{c c c c c c c c c c c c c c c}
\toprule
charge & $s_{AA}$ & $s_{BB}$ & $s_{CC}$ & $s_{DD}$
& $v_{AA}$ & $v_{BB}$ & $v_{CC}$ & $v_{DD}$
& $w_{AB}$ & $w_{AC}$ & $w_{AD}$ & $w_{BC}$ & $w_{BD}$ & $w_{CD}$\\
\midrule
$q_{12}$ & $3/2$ & $3/2$ & $3$ & $4$ & $1/4$ & $-1/4$ & $0$ & $0$ & $1/4$ & $0$ & $0$ & $0$ & $0$ & $0$\\
$-q_{12}$ & $3/2$ & $3/2$ & $3$ & $4$ & $1/4$ & $-1/4$ & $0$ & $0$ & $-1/4$ & $0$ & $0$ & $0$ & $0$ & $0$\\
$q_{13}$ & $2$ & $2$ & $2$ & $4$ & $1/2$ & $0$ & $-1/2$ & $0$ & $0$ & $1/2$ & $0$ & $0$ & $0$ & $0$\\
$-q_{13}$ & $2$ & $2$ & $2$ & $4$ & $1/2$ & $0$ & $-1/2$ & $0$ & $0$ & $-1/2$ & $0$ & $0$ & $0$ & $0$\\
$q_{14}$ & $5/2$ & $2$ & $3$ & $5/2$ & $3/4$ & $0$ & $0$ & $-3/4$ & $0$ & $0$ & $3/4$ & $0$ & $0$ & $0$\\
$-q_{14}$ & $5/2$ & $2$ & $3$ & $5/2$ & $3/4$ & $0$ & $0$ & $-3/4$ & $0$ & $0$ & $-3/4$ & $0$ & $0$ & $0$\\
$q_{23}$ & $1$ & $5/2$ & $5/2$ & $4$ & $0$ & $1/4$ & $-1/4$ & $0$ & $0$ & $0$ & $0$ & $1/4$ & $0$ & $0$\\
$-q_{23}$ & $1$ & $5/2$ & $5/2$ & $4$ & $0$ & $1/4$ & $-1/4$ & $0$ & $0$ & $0$ & $0$ & $-1/4$ & $0$ & $0$\\
$q_{24}$ & $1$ & $3$ & $3$ & $3$ & $0$ & $1/2$ & $0$ & $-1/2$ & $0$ & $0$ & $0$ & $0$ & $1/2$ & $0$\\
$-q_{24}$ & $1$ & $3$ & $3$ & $3$ & $0$ & $1/2$ & $0$ & $-1/2$ & $0$ & $0$ & $0$ & $0$ & $-1/2$ & $0$\\
$q_{34}$ & $1$ & $2$ & $7/2$ & $7/2$ & $0$ & $0$ & $1/4$ & $-1/4$ & $0$ & $0$ & $0$ & $0$ & $0$ & $1/4$\\
$-q_{34}$ & $1$ & $2$ & $7/2$ & $7/2$ & $0$ & $0$ & $1/4$ & $-1/4$ & $0$ & $0$ & $0$ & $0$ & $0$ & $-1/4$\\
$+1$ & $1$ & $2$ & $3$ & $4$ & $0$ & $0$ & $0$ & $0$ & $0$ & $0$ & $0$ & $0$ & $0$ & $0$\\
\bottomrule
\end{tabular}
\endgroup
\end{table}

\begin{table}
\caption{Ideal flat-band parameters for the six decompositions of
$\pm q_{1234}$.}
\label{sm:tab:four-band-decompositions}
\centering
\begingroup
\normalsize
\setlength{\tabcolsep}{3.2pt}
\renewcommand{\arraystretch}{1.12}
\begin{tabular}{c c c c c c c c c c c c c c c}
\toprule
decomposition & $s_{AA}$ & $s_{BB}$ & $s_{CC}$ & $s_{DD}$
& $v_{AA}$ & $v_{BB}$ & $v_{CC}$ & $v_{DD}$
& $w_{AB}$ & $w_{AC}$ & $w_{AD}$ & $w_{BC}$ & $w_{BD}$ & $w_{CD}$\\
\midrule
$q_{12}q_{34}$ & $3/2$ & $3/2$ & $7/2$ & $7/2$ & $1/4$ & $-1/4$ & $1/4$ & $-1/4$ & $1/4$ & $0$ & $0$ & $0$ & $0$ & $1/4$\\
$-q_{12}q_{34}$ & $3/2$ & $3/2$ & $7/2$ & $7/2$ & $1/4$ & $-1/4$ & $1/4$ & $-1/4$ & $-1/4$ & $0$ & $0$ & $0$ & $0$ & $1/4$\\
$-q_{13}q_{24}$ & $2$ & $3$ & $2$ & $3$ & $1/2$ & $1/2$ & $-1/2$ & $-1/2$ & $0$ & $-1/2$ & $0$ & $0$ & $1/2$ & $0$\\
$q_{13}q_{24}$ & $2$ & $3$ & $2$ & $3$ & $1/2$ & $1/2$ & $-1/2$ & $-1/2$ & $0$ & $1/2$ & $0$ & $0$ & $1/2$ & $0$\\
$q_{14}q_{23}$ & $5/2$ & $5/2$ & $5/2$ & $5/2$ & $3/4$ & $1/4$ & $-1/4$ & $-3/4$ & $0$ & $0$ & $3/4$ & $1/4$ & $0$ & $0$\\
$-q_{14}q_{23}$ & $5/2$ & $5/2$ & $5/2$ & $5/2$ & $3/4$ & $1/4$ & $-1/4$ & $-3/4$ & $0$ & $0$ & $-3/4$ & $1/4$ & $0$ & $0$\\
\bottomrule
\end{tabular}
\endgroup
\end{table}

For the charge $-1=q_{12}^2$, we use a flat-band model with
$(s_{AA},s_{BB},s_{CC},s_{DD})=(3/2,3/2,3,4)$,
$v'_{AA}=-1/4$, $v'_{BB}=1/4$, and $w'_{AB}=-1/4$;
all other hopping coefficients vanish.

{

\subsection{Gap-resolved quench dynamics}
\label{sm:sec:four-band-overview}

Following the three-band protocol, we perform separate quenches initialized
on the four sublattice sites within a bulk unit cell, under the same
post-quench Hamiltonian. The time-evolved states are assembled into the
full evolution operator $U(q,t)$.
For each gap, we collect the bands below and above it into two groups
and retain their relative dynamics through band projection. This gives
three gap-resolved evolution operators $U_g(q,t)$, with $g=1,2,3$,
and the corresponding Bloch vectors $\bm n_g(q,t)$. The projections
and their relation to effective two-band models with chiral symmetry
are derived in Secs.~\ref{sm:sec:four-band-reduction}
and~\ref{sm:sec:four-band-chiral-topology}. The gap-resolved Chern numbers
$(C_1,C_2,C_3)$ are then calculated over the emergent
quasimomentum--time manifolds, as in the three-band case.

A skyrmion in gap 1, 2, or 3 carries the charge $\pm q_{12}$,
$\pm q_{23}$, or $\pm q_{34}$, respectively. These elementary charges
generate $Q_{16}$, so any total topological charge can be expressed
as their ordered product, with the multiplication rules given in
Sec.~\ref{sm:sec:q16}. For the models considered here, the magnitudes
$|C_g|$ count the occurrences of the corresponding generators in gap $g$
and predict the edge-mode patterns, while the signs of $C_g$ distinguish
the directions of the eigenframe rotations. The global charge can
then be determined from $(C_1,C_2,C_3)$, except that distinguishing
$q_{1234}$ from $-q_{1234}$ can require an additional quench measurement.

\begin{figure}
\centering
\includegraphics[width=0.5\textwidth]{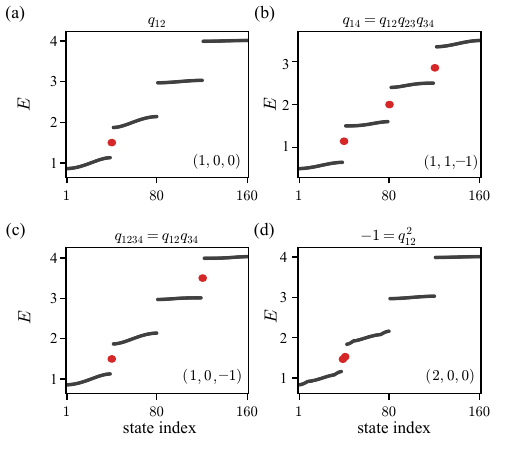}
\caption{Open-boundary spectra and gap-resolved Chern numbers
for the four-band model. Black and red dots denote bulk and edge states,
respectively. The gap-resolved Chern numbers $(C_1,C_2,C_3)$ are shown
in each panel. The labels above the panels indicate the global
topological charges and their decompositions:
(a) $q_{12}$, (b) $q_{14}=q_{12}q_{23}q_{34}$,
(c) $q_{1234}=q_{12}q_{34}$, and (d) $-1=q_{12}^2$.}
\label{sm:fig:four-band-spectra}
\end{figure}

Figure~\ref{sm:fig:four-band-spectra} shows the open-boundary spectra
for four representative models, together with their gap-resolved
Chern numbers. In panel (a),
$(C_1,C_2,C_3)=(1,0,0)$ is consistent with the charge $q_{12}$ and a
pair of edge modes in gap 1. In panel (b), $(C_1,C_2,C_3)=(1,1,-1)$
corresponds to $q_{14}=q_{12}q_{23}q_{34}$, with a pair of edge modes
in each gap. Panel (c) shows the realization $q_{1234}=q_{12}q_{34}$,
with $(C_1,C_2,C_3)=(1,0,-1)$ and a pair of edge modes in gaps 1 and 3.
Finally, panel (d) shows $-1=q_{12}^2$, with
$(C_1,C_2,C_3)=(2,0,0)$ and two pairs of edge modes in gap 1.

Figure~\ref{sm:fig:four-band-bloch-vectors} shows the gap-resolved spin
textures of $\bm n_g(q,t)$ for six four-band models. The three gaps in
each model are shown in the same column, with all three textures
calculated from the same evolution operator.

\begin{figure*}
\centering
\includegraphics[width=\textwidth]{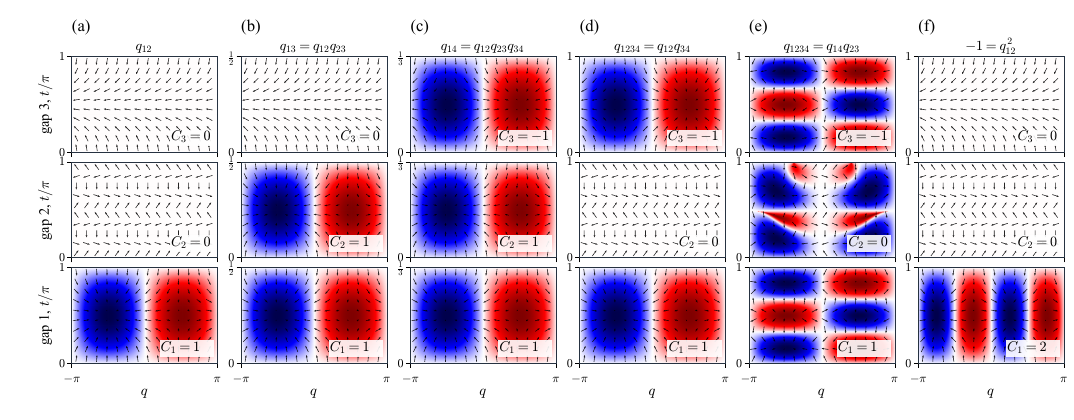}
\caption{Gap-resolved spin textures for
(a) $q_{12}$, (b) $q_{13}=q_{12}q_{23}$,
(c) $q_{14}=q_{12}q_{23}q_{34}$,
(d) $q_{1234}=q_{12}q_{34}$,
(e) $q_{1234}=q_{14}q_{23}$, and (f) $-1=q_{12}^2$.
Within each column,
gap 1, gap 2, and gap 3 are shown from bottom to top. The background color
denotes $n_{g,z}$ from $-1$ to $+1$,
and black arrows show $(n_{g,x},n_{g,y})$. Each panel is labeled by the
corresponding $C_g$.}
\label{sm:fig:four-band-bloch-vectors}
\end{figure*}

The need for an additional quench arises because different eigenframe
rotations can give the same gap-resolved Chern numbers. For example,
$-q_{12}q_{34}=-q_{1234}$ and $-q_{13}q_{24}=+q_{1234}$ both give
$(C_1,C_2,C_3)=(-1,0,-1)$ under our sign convention. For these models, a quench
initialized on sublattice $A$ distinguishes the two cases through the
populations transferred to sublattices $B$ and $C$. More generally,
measuring the populations on $B$, $C$, and $D$ distinguishes the three
decompositions of $\pm q_{1234}$ into rotations in orthogonal planes,
as detailed in Sec.~\ref{sm:sec:q16-skyrmion-correspondence}.
Together with the gap-resolved Chern numbers, this measurement determines
the global non-Abelian charge for the models considered here.

}

\subsection{\texorpdfstring{$Q_{16}$}{Q16} algebra and multiplication table}
\label{sm:sec:q16}

The charges $q_{ij}$ satisfy $q_{ij}^2=-1$.
Two such charges anticommute if they share exactly one band index,
and commute if their band pairs are disjoint~\cite{jiang2021fourband}.
The charge $q_{1234}$ has three decompositions into rotations in
orthogonal planes,
\begin{equation}
q_{1234}=q_{12}q_{34}=-q_{13}q_{24}=q_{14}q_{23}.
\label{sm:eq:q1234-decompositions}
\end{equation}

\begin{table}
\caption{Products of the eight representatives without an overall minus
sign.  The entry in row $g$ and column $h$ is the ordered product $gh$; a
minus sign in an entry is part of that product.}
\label{sm:tab:q16}
\centering
\begingroup
\small
\setlength{\tabcolsep}{5.0pt}
\renewcommand{\arraystretch}{1.18}
\begin{tabular}{c|*{8}{c}}
$\times$ & $1$ & $q_{12}$ & $q_{13}$ & $q_{14}$ & $q_{23}$ & $q_{24}$ & $q_{34}$ & $q_{1234}$\\
\hline
$1$ & $1$ & $q_{12}$ & $q_{13}$ & $q_{14}$ & $q_{23}$ & $q_{24}$ & $q_{34}$ & $q_{1234}$ \\
$q_{12}$ & $q_{12}$ & $-1$ & $q_{23}$ & $q_{24}$ & $-q_{13}$ & $-q_{14}$ & $q_{1234}$ & $-q_{34}$ \\
$q_{13}$ & $q_{13}$ & $-q_{23}$ & $-1$ & $q_{34}$ & $q_{12}$ & $-q_{1234}$ & $-q_{14}$ & $q_{24}$ \\
$q_{14}$ & $q_{14}$ & $-q_{24}$ & $-q_{34}$ & $-1$ & $q_{1234}$ & $q_{12}$ & $q_{13}$ & $-q_{23}$ \\
$q_{23}$ & $q_{23}$ & $q_{13}$ & $-q_{12}$ & $q_{1234}$ & $-1$ & $q_{34}$ & $-q_{24}$ & $-q_{14}$ \\
$q_{24}$ & $q_{24}$ & $q_{14}$ & $-q_{1234}$ & $-q_{12}$ & $-q_{34}$ & $-1$ & $q_{23}$ & $q_{13}$ \\
$q_{34}$ & $q_{34}$ & $q_{1234}$ & $q_{14}$ & $-q_{13}$ & $q_{24}$ & $-q_{23}$ & $-1$ & $-q_{12}$ \\
$q_{1234}$ & $q_{1234}$ & $-q_{34}$ & $q_{24}$ & $-q_{23}$ & $-q_{14}$ & $q_{13}$ & $-q_{12}$ & $1$ \\
\end{tabular}%
\endgroup
\end{table}

If $g$ and $h$ are representatives in Table~\ref{sm:tab:q16} and
$s,t=\pm1$, all products involving an overall sign follow from
$(sg)(th)=st(gh)$.  Thus the reduced table fixes the multiplication of all
sixteen elements without repeating sign-related rows and columns.

\subsection{Projection from four-band evolution to effective two-band evolution}
\label{sm:sec:four-band-reduction}

The three physical gaps correspond to the partitions
\begin{equation}
\mathrm{gap}\,1:1|234,\qquad
\mathrm{gap}\,2:12|34,\qquad
\mathrm{gap}\,3:123|4.
\label{sm:eq:four-band-partitions}
\end{equation}
We choose a reference quasimomentum $q_0$ and denote the corresponding
eigenframe by
\begin{equation}
|e_i\rangle\equiv |u_i(q_0)\rangle,
\qquad i=1,2,3,4.
\label{sm:eq:four-band-reference-frame}
\end{equation}
The vectors $\{|e_i\rangle\}$ form the fixed reference basis used in the
gap-resolved projections below.
The outer gaps contain a distinguished eigenline on one side and can be
treated by the same line/composite construction as the three-band problem.
For gap 1, let $P_{234}=I-|e_1\rangle\langle e_1|$ and define
\begin{equation}
W_1(q)=\left(
|e_1\rangle,
\chi_1(q)\frac{P_{234}|u_1(q)\rangle}{\|P_{234}|u_1(q)\rangle\|}
\right),\qquad
U_1(q,t)=W_1^\dagger(q)U(q,t)W_1(q).
\label{sm:eq:four-band-gap1}
\end{equation}
For gap 3, with $P_{123}=I-|e_4\rangle\langle e_4|$,
\begin{equation}
W_3(q)=\left(
|e_4\rangle,
\chi_3(q)\frac{P_{123}|u_4(q)\rangle}{\|P_{123}|u_4(q)\rangle\|}
\right),\qquad
U_3(q,t)=W_3^\dagger(q)U(q,t)W_3(q).
\label{sm:eq:four-band-gap3}
\end{equation}
The signs $\chi_{1,3}(q)=\pm1$ fix the continuous orientation of the
second basis vector, as in the three-band construction.

The middle gap is qualitatively different because both sides of the
partition $12|34$ are two-dimensional. There is therefore no distinguished
eigenline that can be used directly in the composite construction
above. Choosing an individual state inside either two-band sector would be
arbitrary, since such a state changes under an internal $O(2)$ rotation of
the corresponding pair of eigenvectors.

To remove this internal basis dependence, we represent the lower two-band
subspace $\mathrm{span}\{|u_1(q)\rangle,|u_2(q)\rangle\}$ by the
decomposable two-vector
\begin{equation}
|u_{12}(q)\rangle
=
|u_1(q)\rangle\wedge|u_2(q)\rangle
\in\Lambda^2\mathbb R^4 .
\label{sm:eq:plucker-line}
\end{equation}
Here $\Lambda^2\mathbb R^4$ is the six-dimensional exterior-product space~\cite{bouhon2020geometry}.
Using the reference eigenframe
$|e_a\rangle$, we define
\begin{equation}
|e_{ab}\rangle
\equiv
|e_a\rangle\wedge|e_b\rangle,
\qquad 1\leq a<b\leq4,
\label{sm:eq:exterior-basis}
\end{equation}
so that
\begin{equation}
\{
|e_{12}\rangle,
|e_{13}\rangle,
|e_{14}\rangle,
|e_{23}\rangle,
|e_{24}\rangle,
|e_{34}\rangle
\}
\end{equation}
forms an orthonormal basis of $\Lambda^2\mathbb R^4$.
The exterior-product state has the expansion
\begin{equation}
|u_{12}(q)\rangle
=
\sum_{a<b}p_{ab}(q)|e_{ab}\rangle ,
\label{sm:eq:plucker-expansion}
\end{equation}
where the Pl\"ucker coordinates are
\begin{equation}
p_{ab}(q)=u_{1a}(q)u_{2b}(q)-u_{1b}(q)u_{2a}(q),
\qquad a<b.
\label{sm:eq:plucker-coordinates}
\end{equation}
Here $u_{na}(q)=\langle e_a|u_n(q)\rangle$.

The exterior-product state $|u_{12}(q)\rangle$ depends only on the
two-dimensional subspace spanned by $|u_1(q)\rangle$ and $|u_2(q)\rangle$,
rather than on the particular orthonormal basis chosen within that subspace,
up to an overall sign. Thus the lower two-band subspace is represented by
a real line in $\Lambda^2\mathbb R^4$, allowing the middle-gap problem to
be treated in the same way as the outer gaps.

We can then apply the same two-dimensional projection in
$\Lambda^2\mathbb R^4$. Taking $|e_{12}\rangle$ as the fixed reference
direction, we define
\begin{equation}
P_{12}^{\perp}
=
I-|e_{12}\rangle\langle e_{12}|,
\end{equation}
where $I$ is the identity operator in $\Lambda^2\mathbb R^4$. The
corresponding two-dimensional basis is
\begin{equation}
\mathcal W_2(q)
=
\left(
|e_{12}\rangle,\,
\chi_2(q)
\frac{
P_{12}^{\perp}|u_{12}(q)\rangle
}{
\|P_{12}^{\perp}|u_{12}(q)\rangle\|
}
\right).
\label{sm:eq:plucker-basis}
\end{equation}
The sign $\chi_2(q)=\pm1$ is chosen according to the same orientation
convention as for the outer gaps.
The effective middle-gap evolution is defined as
\begin{equation}
U_2(q,t)
=
\mathcal W_2^\dagger(q)
[\Lambda^2U(q,t)]
\mathcal W_2(q).
\label{sm:eq:four-band-gap2}
\end{equation}
Here $\Lambda^2U(q,t)$ denotes the evolution induced by $U(q,t)$ in
$\Lambda^2\mathbb R^4$, defined through
\begin{equation}
(\Lambda^2U)
\bigl(|v\rangle\wedge|w\rangle\bigr)
=
U|v\rangle\wedge U|w\rangle .
\label{sm:eq:exterior-action}
\end{equation}
Equations~\eqref{sm:eq:four-band-gap1}, \eqref{sm:eq:four-band-gap3},
and~\eqref{sm:eq:four-band-gap2} give the projected evolution operators
for the three gaps. The normalized Bloch vectors are then obtained from
Eq.~(5) of the main text.

\subsection{Effective chiral two-band dynamics}
\label{sm:sec:four-band-chiral-topology}

As in the three-band case, the connection to an effective two-band
Hamiltonian with chiral symmetry can be understood by flattening the
energies on the two sides of the selected gap while keeping the
corresponding band subspaces unchanged. We give the construction for a
PT-symmetric $n$-band Hamiltonian, which includes the four-band case.

Consider the gap between bands $g$ and $g+1$. Keeping all eigenstates fixed,
we continuously move the energies below this gap to $E_-$ and those above
it to $E_+$, with $E_-<E_+$. The selected gap remains open throughout this
deformation, giving
\begin{equation}
H_g^{\rm flat}(q)
=
E_-P_g(q)+E_+\bigl[I_n-P_g(q)\bigr],
\qquad
P_g(q)=\sum_{a=1}^g|u_a(q)\rangle\langle u_a(q)|.
\label{sm:eq:n-band-flat-hamiltonian}
\end{equation}
Since the eigenstates are unchanged, the $g$-dimensional band subspace
associated with the selected gap is unchanged by the flattening. We encode
this subspace by
\begin{equation}
|p_g(q)\rangle
=
|u_1(q)\rangle\wedge\cdots\wedge|u_g(q)\rangle
\in\Lambda^g\mathbb R^n .
\label{sm:eq:n-band-plucker}
\end{equation}
As in the four-band construction above, $|p_g(q)\rangle$ represents the
entire $g$-dimensional subspace independently of the particular orthonormal
basis chosen within it, up to an overall sign. Thus the subspace is mapped
to a real line in $\Lambda^g\mathbb R^n$.

For the flattened Hamiltonian, this line is an eigenspace of the induced
Hamiltonian in $\Lambda^g\mathbb R^n$ with eigenvalue $gE_-$. Any state
containing at least one eigenvector from above the selected gap is separated
from it by at least $E_+-E_-$. We may therefore further flatten the spectrum
in the exterior-product space while keeping $|p_g(q)\rangle$ fixed, giving
\begin{equation}
\mathcal H_g^{\rm flat}(q)
=
\varepsilon_-|p_g(q)\rangle\langle p_g(q)|
+\varepsilon_+\bigl[I-|p_g(q)\rangle\langle p_g(q)|\bigr],
\qquad
\varepsilon_-<\varepsilon_+.
\label{sm:eq:n-band-flat-generator}
\end{equation}

We then construct the two-dimensional basis $\mathcal W_g(q)$ from a fixed
reference direction and the normalized perpendicular component of
$|p_g(q)\rangle$, following the same procedure as above. Writing
\begin{equation}
\mathcal W_g^\dagger|p_g\rangle
=
\begin{pmatrix}
a_g\\
b_g
\end{pmatrix},
\qquad
a_g^2+b_g^2=1,
\end{equation}
the projected Hamiltonian, after subtracting its mean energy, becomes
\begin{equation}
\widetilde h_g^{\rm flat}(q)
=
\mathcal W_g^\dagger\mathcal H_g^{\rm flat}\mathcal W_g
-\bar\varepsilon I_2
=
\frac{\Delta}{2}
\left[
2a_gb_g\sigma_x
+
(a_g^2-b_g^2)\sigma_z
\right],
\qquad
\bar\varepsilon=\frac{\varepsilon_-+\varepsilon_+}{2},
\quad
\Delta=\varepsilon_--\varepsilon_+.
\label{sm:eq:n-band-chiral-reduction}
\end{equation}
Only $\sigma_x$ and $\sigma_z$ appear, and therefore
\begin{equation}
\{\widetilde h_g^{\rm flat},\sigma_y\}=0.
\end{equation}
The effective two-band Hamiltonian thus has chiral symmetry, showing that
each physical gap can be reduced to the same chiral two-band form used in
the gap-resolved quench construction.

\subsection{Correspondence between the Chern numbers and topological charges}
\label{sm:sec:q16-skyrmion-correspondence}

Table~\ref{sm:tab:q16-skyrmion} summarizes the correspondence between
the gap-resolved Chern numbers $(C_1,C_2,C_3)$ and the topological
charges for the models considered here, excluding $\pm q_{1234}$.
The signs follow Eq.(6) of the main text,
including its overall minus sign, and agree with the labels in
Fig.~\ref{sm:fig:four-band-bloch-vectors}. The same global charge can
be realized by different sequences of eigenframe rotations, so these
numbers refer to the particular models specified above.

\begin{table}
\centering
\caption{Gap-resolved Chern numbers for representative
$Q_{16}$ charges, excluding the $\pm q_{1234}$ sector, under the
convention used here.}
\label{sm:tab:q16-skyrmion}
\begin{tabular}{cc}
\toprule
Charge & $(C_1,C_2,C_3)$\\
\midrule
$+1$           & $(0,0,0)$\\
$-1$           & $(2,0,0)$\\
$\pm q_{12}$   & $(\pm 1,0,0)$\\
$\pm q_{23}$   & $(0,\pm1,0)$\\
$\pm q_{34}$   & $(0,0,\mp 1)$\\
$\pm q_{13}$   & $(\pm1,\pm1,0)$\\
$\pm q_{24}$   & $(0,\pm1,\mp1)$\\
$\pm q_{14}$   & $(\pm1,\pm1,\mp1)$\\
\bottomrule
\end{tabular}
\end{table}

The $-1$ entry in Table~\ref{sm:tab:q16-skyrmion} corresponds to the
explicit double-winding representative $q_{12}^{2}=-1$, for which the
two elementary contributions occur in gap 1. Equivalent
representatives constructed from double windings in other rotation
planes can instead produce $C_g=\pm2$ in the corresponding gaps.
Thus, the gap-resolved Chern numbers retain information about how the elementary
rotations are distributed among the physical gaps in addition to the
global $Q_{16}$ charge.

The charges $\pm q_{1234}$ each admits three minimal decompositions
into rotations in mutually orthogonal planes.
Table~\ref{sm:tab:q1234-six-representatives} gives the Chern numbers
for the six models in Table~\ref{sm:tab:four-band-decompositions}.
The last column specifies the sublattice $X_*$ with the largest
transferred population after a quench initialized on sublattice $A$,
as defined below.

\begin{table}
\centering
\caption{Six representatives of $\pm q_{1234}$, their
gap-resolved Chern numbers, and the dominant transfer channel
following a quench initialized on sublattice $A$.}
\label{sm:tab:q1234-six-representatives}
\begingroup
\small
\setlength{\tabcolsep}{4pt}
\begin{tabular}{cccc}
\toprule
Charge & decomposition & $(C_1,C_2,C_3)$ & $X_*$\\
\midrule
$+q_{1234}$ & $q_{12}q_{34}$   & $(+1,0,-1)$ & $B$\\
$-q_{1234}$ & $-q_{12}q_{34}$  & $(-1,0,-1)$ & $B$\\
$+q_{1234}$ & $-q_{13}q_{24}$  & $(-1,0,-1)$ & $C$\\
$-q_{1234}$ & $q_{13}q_{24}$   & $(+1,0,-1)$ & $C$\\
$+q_{1234}$ & $q_{14}q_{23}$   & $(+1,0,-1)$ & $D$\\
$-q_{1234}$ & $-q_{14}q_{23}$  & $(-1,0,+1)$ & $D$\\
\bottomrule
\end{tabular}
\endgroup
\end{table}

The factors $\chi_g$ fix the rotation sense within each individual
gap-resolved reduction, so that the Chern numbers retain the
orientation information of the corresponding effective two-band
evolutions. The remaining ambiguity in the $\pm q_{1234}$ sector has
a different origin: distinct decompositions into two orthogonal
rotation planes can yield the same gap-resolved Chern numbers
$(C_1,C_2,C_3)$. For example, $(-1,0,-1)$ occurs for both
$-q_{12}q_{34}=-q_{1234}$ and
$-q_{13}q_{24}=+q_{1234}$. Likewise, $(+1,0,-1)$ occurs for three
representatives, two carrying $+q_{1234}$ and one carrying
$-q_{1234}$. Therefore, although the Chern numbers resolve
the orientation within each gap reduction, they do not uniquely
determine the decomposition of the full four-band eigenframe
evolution in the $\pm q_{1234}$ sector.

To resolve this ambiguity, we perform an additional quench under the
same bulk Hamiltonian. We prepare
a state $|R_0,A\rangle$ localized on sublattice $A$ in a bulk unit
cell and measure the total population transferred to each of the
other three sublattices. For a translationally invariant bulk,
the transferred population is

\begin{equation}
P_{A\rightarrow X}(t)
=
\sum_R
\left|
\langle R,X|e^{-iHt}|R_0,A\rangle
\right|^2
=
\int_{-\pi}^{\pi}\frac{dq}{2\pi}
\left|U_{XA}(q,t)\right|^2,
\qquad X=B,C,D .
\label{sm:eq:q1234-sublattice-transfer}
\end{equation}
We choose a readout time $t_r$ at which the representative models
show a nonvanishing transfer contrast and define
\begin{equation}
X_*=
\operatorname*{arg\,max}_{X=B,C,D}
P_{A\rightarrow X}(t_r).
\end{equation}
For the three decompositions into rotations in orthogonal
planes, the rotation plane
containing the initial $A$ component is respectively $AB$, $AC$, and
$AD$ for the $12|34$, $13|24$, and $14|23$ pairings. The dominant
transfer channel therefore provides an auxiliary label that
distinguishes these three decompositions.

Combining this auxiliary label with the gap-resolved Chern numbers
resolves the ambiguity among the six representatives. For
$(-1,0,-1)$, $X_*=B$ corresponds to
$-q_{12}q_{34}=-q_{1234}$, whereas $X_*=C$ corresponds to
$-q_{13}q_{24}=+q_{1234}$. For $(+1,0,-1)$, the outcomes
$X_*=B,C,D$ correspond respectively to
$q_{12}q_{34}=+q_{1234}$,
$q_{13}q_{24}=-q_{1234}$, and
$q_{14}q_{23}=+q_{1234}$. The remaining set of Chern numbers $(-1,0,+1)$ already
selects the $-q_{14}q_{23}=-q_{1234}$ representative
among these six models.

Combining the gap-resolved Chern numbers with the populations measured
in this additional quench allows all sixteen elements of $Q_{16}$ to be
distinguished for the models considered here.

\end{document}